\documentclass[twocolumn]{pazhb_eng}

\usepackage{graphicx}
\usepackage{amsmath}
\usepackage{rotating}
\usepackage{lscape}
\usepackage{ifpdf}
\usepackage{breqn}
\usepackage[hyphens]{url}

\newcommand {\be}{\begin{equation}}
\newcommand {\ee}{\end{equation}}

\newcommand{\angstrom}{\mbox{\normalfont\AA}}

\begin{document}
\journalinfo{2018}{44}{8-9}{522}[540]

\title{Optical Identification of X-ray Sources from the 14-Year INTEGRAL All-Sky Survey
}
\author{\bf D.I. Karasev$^1$\email{dkarasev@iki.rssi.ru}, A.A.Lutovinov$^1$, A. Yu. Tkachenko$^1$, G. A. Khorunzhev$^1$, R. A. Krivonos$^1$, P. S. Medvedev$^1$, M. N. Pavlinsky$^1$, R. A. Burenin$^1$ and M. V. Eselevich$^2$\\
$^1$\it{Space Research Institute, Moscow, Russia\\}
$^2$\it{Institute of Solar–Terrestrial Physics, Russian Academy of Sciences, Siberian Branch, P.O. Box 4026, Irkutsk, 664033 Russia}
}
\shortauthor{}
\shorttitle{}
\submitted{6 March 2018}

\begin{abstract}
We present the results of our optical identifications of several hard X-ray sources from the INTEGRAL all-sky survey obtained over 14 years of observations. Having improved the positions of these objects in the sky with the X-ray telescope (XRT) of the Swift observatory and the XMM-Newton observatory, we have identified their counterparts using optical and infrared sky survey data. We have obtained optical spectra for more than half of the objects from our sample with the RTT-150 and AZT-33IK telescopes, which have allowed us to establish the nature of the objects and to measure their redshifts. Six sources are shown to be extragalactic in origin and to belong to Seyfert 1 and 2 galaxies (IGR\,J01017+6519, IGR\,J08215-1320, IGR\,J08321-1808, IGR\,J16494-1740, IGR\,J17098- 2344, IGR\,J17422-2108); we have failed to draw definitive conclusions about the nature of four more objects (IGR\,J11299-6557, IGR\,J14417-5533, IGR\,J18141-1823, IGR\,J18544+0839), but, judging by circumstantial evidence, they are most likely also extragalactic objects. For one more object (IGR\,J18044-1829) no unambiguous identification has been made.

\medskip
\end{abstract}

\section{INTRODUCTION}

The INTEGRAL observatory (Winkler et al. 2003) has operated in orbit and performed sky observations in the hard X-ray energy range ($>$20 keV) for more than 15 years. In this period a significant exposure has been accumulated and a high sensitivity has been achieved in various sky regions, including the Galactic plane (Revnivtsev et al. 2004, 2006; Molkov et al. 2004; Krivonos et al. 2012, 2017; Bird et al. 2016) and a number of extragalactic fields (Grebenev et al. 2013; Mereminskiy et al. 2016), which has led to the discovery of several hundred new hard X-ray sources (see, e.g., Krivonos et al. 2007, 2012; Bird et al. 2016). In particular, a list of 72 new, previously unknown objects was presented in a recent paper by Krivonos et al. (2017).

The completeness of the catalog of hard X-ray sources detected by the INTEGRAL observatory is very high owing to a large number of works on their identifications in the soft X-ray, visible, and infrared bands (see, e.g., Masetti et al. 2007, 2010; Tomsick et al. 2009, 2016; Malizia et al. 2010). In this paper, using the previously accumulated experience (see, e.g., Bikmaev et al. 2006, 2008; Burenin et al. 2008, 2009; Lutovinov et al. 2010; Karasev et al. 2010), sky survey data in various wavelength ranges, and additional spectroscopic observations with the RTT-150 and AZT-33IK optical telescopes, we investigated 11 objects: IGR\,J01017+6519, IGR\,J08215-1320, IGR\,J08321-1808, IGR\,J11299-6557 IGR\,J14417-5533, IGR\,J16494-1740, IGR\,J17098-2344, IGR\,J17422-2108, IGR\,J18044-1829, IGR\,J18141-1823, and IGR\,J18544+0839 from Krivonos et al. (2017), to determine their nature.

\section{INSTRUMENTS AND DATA}

Coded-aperture instruments allow the positions of sources to be determined with an accuracy that, as a rule, is insufficient for their unequivocal optical identification. Therefore, to improve the localization of the objects from our sample, we used publicly accessible sky observations by the Swift and XMM-Newton space observatories. Since most of the objects being investigated have been detected for the first time and have not been observed previously by other observatories, we performed their additional soft X-ray observations with the X-ray telescope (XRT) of the Neil Gehrels Swift Observatory (hereafter simply the Swift observatory) to improve the localization accuracy.

In most cases we were able to determine the soft X-ray counterparts of the hard X-ray sources detected by the INTEGRAL observatory and to improve their positions in the sky. This allowed the optical object associated with the X-ray source to be determined.

Note that for all objects, based on the significance of their detection, we took the INTEGRAL position accuracy to be $4'$ (Krivonos et al. 2007, 2017). This is a fairly large uncertainty, especially for the Galactic plane; therefore, we ran into ambiguity in choosing a soft X-ray counterpart of the hard X-ray source. The spectra of the hard X-ray sources used here were reconstructed from the INTEGRAL observations over 14 years using special software de- veloped at the Space Research Institute of the Russian Academy of Sciences (see Churazov et al. 2005, 2014; Krivonos et al. 2010).

The XRT/Swift data were processed with the corresponding software\footnote{http://swift.gsfc.nasa.gov} of the HEASOFT 6.22 package\footnote{https://heasarc.nasa.gov/lheasoft/}. The positions of the objects in the XRT/Swift images and their localization accuracy were determined with standard recommended procedures and
algorithms\footnote{http://www.swift.ac.uk/user\_objects/} (Goad et al. 2007; Evans et al. 2009).
The XMM-Newton data (archival observations were available for two objects from the sample) were processed with the SAS version 16.1 software and the most up-to-date calibration data. In what follows, we will provide the sky images obtained by the X-ray telescopes only if there is ambiguity in choosing a soft X-ray counterpart of the hard X-ray source. The X-ray spectra of the sources were fitted using the XSPEC package.

Since the observations of the soft and hard (INTEGRAL data) X-ray sources occurred at different times, in our joint analysis of the spectra we added an additional parameter (floating normalization) to the model to take into account the possible variability of the X-ray flux from the sources. This parameter is measured relative to the INTEGRAL data and corresponds to the coefficient by which the data from other observatories should be divided to equalize the normalizations. Among other things, we provide this parameter in the final table. When simultaneously modeling the spectra, we used the $\chi^2$-statistic; to determine the X-ray flux and luminosity in the soft X-ray band, we modeled only the soft X-ray spectra and used the $C$-statistic (cstat).

The positions of the sources in the optical and near-infrared bands presented in the paper were taken from the publicly accessible catalogs of the PanSTARRS, UKIDSS, and VVV sky surveys. We investigated the objects in the mid-infrared based on data from the open source catalogs of the Spitzer (GLIMPSE survey) and WISE space observatories.

To perform spectroscopic studies of some of the objects from the sample, we used the observational data from the Russian-Turkish 1.5-m telescope (RTT-150) obtained with the medium- and low- resolution TFOSC spectrometer\emph{TFOSC}\footnote{http://hea.iki.rssi.ru/rtt150/ru/index.php?page=tfosc}. For the spectroscopy we used the N15 grism that gives the widest wavelength range ($3500-9000 \angstrom$)and the greatest quantum efficiency. The spectral resolution in this case was $\approx12 \angstrom$ (full width at half maximum).
Apart from RTT-150, for our spectroscopic observations we also used the 1.6-m AZT-33IK telescope of the Sayan observatory (Kamus et al. 2002), the ADAM spectrograph with a 2-arcsec slit and a VPHG600R grating (Afanasiev et al. 2016; Burenin et al. 2016). The spectral resolution of the instrument
is $\approx7.5\angstrom$ (full width at half maximum) in the wavelength range $6520 - 10100 \angstrom$.

All of the spectroscopic observations were proessed in a standard way using the IRAF\emph{IRAF}\footnote{http://iraf.noao.edu} software and our own software package. To calculate the photometric distance $D_{L}$, we used the cosmological parameters $H_0$= 67.8 (km/s)/Mpc and $\Omega_M$ = 0.308 (Planck Collaboration 2016).

\section{OPTICAL IDENTIFICATION OF SOURCES}
Basic data on the sources being investigated in this paper are presented in Table 1. It provides the names of the sources, the coordinates of their putative soft X-ray counterparts from the Swift and XMM-Newton data, the localization accuracy, and the total flux.
Detailed information about the properties and presumed nature of each of the objects listed in Table 1 is given below.

\bigskip

\begin{table*}[h]
\centering
\footnotesize{
   \caption{Soft X-ray counterparts of the sources being investigated}
   \begin{tabular}{c|c|c|c|c|c}
     \hline
     \hline
           &     &     & Localization   & Flux (2–10 keV) & \\
          Name    & RA(J2000) & Dec(J2000) & accuracy  & $\times10^{-12}$  &Notes\\
              &         &         & (90\%), arcsec      & erg s$^{-1}$ cm$^{-2} $&\\

     \hline

     IGR\,J01017+6519$^*$    &  01$^h$ 01$^m$ 58$^s$.15 &  $+65$\fdg$ 21$\arcmin$ 19$\arcsec$.2$ & 2.6 & ${1.91\pm0.47}$ & - \\
     IGR\,J08215-1320$^*$    &  08$^h$ 21$^m$ 33$^s$.61 &  $-13$\fdg$ 21$\arcmin$ 03$\arcsec$.9$ & 5.3 & $ 0.31\pm0.06$ &  MCG-02-22-003\\
     IGR\,J08321-1808    &  08$^h$ 31$^m$ 58$^s$.40 &  $-18$\fdg$ 08$\arcmin$ 36$\arcsec$.8$ & 2.6 & ${2.35\pm0.61}$ & 1RXSJ083158.1-180828 \\
         &   &   &  & & 1SXPS J083158.6-180840 \\
         
         IGR\,J11299-6557$^*$ &  11$^h$ 29$^m$ 56$^s$.63 & $-65$\fdg$ 55$\arcmin$ 21$\arcsec$.4$ & 2.3 & ${2.03\pm0.51}$ & - \\

     IGR\,J14417-5533$^*$ &  14$^h$ 41$^m$ 18.$^s$.67 & $-55$\fdg$ 33$\arcmin$ 35$\arcsec$.7$ & 2.2 & ${3.78\pm0.49}$ &  \\
          
     IGR\,J16494-1740    &  16$^h$ 49$^m$ 21$^s$.00 &  $-17$\fdg$ 38$\arcmin$ 41$\arcsec$.1$ & 2.3 & ${4.16\pm0.72}$ & 1SXPS J164920.9-173840 \\
     IGR\,J17098-2344$^{X}$	 &  17$^h$ 09$^m$ 44$^s$.70 &  $-23$\fdg$ 46$\arcmin$ 53$\arcsec$.6$ & 2 & ${4.72\pm0.21}$ & 1RXS J170944.9-234658\\	
     IGR\,J17422-2108$^{X}$	&  17$^h$ 42$^m$ 11$^s$.40 &  $-21$\fdg$ 03$\arcmin$ 53$\arcsec$.2$ & 2 & ${2.51\pm0.75}$ & 1SXPS J174211.7-210354 \\	
     
     IGR\,J18044-1829$^*$ &  18$^h$ 04$^m$ 33$^s$.91 & $-18$\fdg$ 30$\arcmin$ 08$\arcsec$.5$ & 2.6 & ${2.39\pm0.74}$ & - \\			
     					
     IGR\,J18141-1823 &  18$^h$ 14$^m$ 14$^s$.54 & $-18$\fdg$ 23$\arcmin$ 11$\arcsec$.4$ & 3.6 &${2.11\pm0.45}$  & 1SXPS J181414.8-182310\\

     IGR\,J18544+0839$^*$ &  18$^h$ 54$^m$ 22$^s$.37 & $+08$\fdg$ 38$\arcmin$ 46$\arcsec$.7$ & 2.3 & ${2.70\pm0.62}$ & - \\

     \hline

     \hline

    \end{tabular}
    *- Obtained from the XRT/Swift observations requested by our group. 
     
    $^{X}$- Based on the archival XMM-Newton data.
    }

\end{table*}

\subsection{IGR\,J01017+6519}

Since previously the sky region around the source was not observed in the soft X-ray band, we requested the XRT/Swift observations (ObsID 00010021001, April 2017). Based on these observations, we were able to localize the source with a good accuracy, which allowed its counterpart in the optical and infrared bands to be identified (Table 1, Fig. 1). The broadband spectrum of the source is well described by a power law with absorption (Fig. 1, Table 2).

To identify the source in the optical and infrared bands, we used the PanSTARRS and WISE data. The PanSTARRS data were used to determine the object’s position with a high accuracy (Table 3), while owing to the WISE data we were able to draw preliminary conclusions about its nature. In particular, it follows from the mid-infrared WISE photometric data that the star-like object falling into the XRT/Swift error circle has a noticeable infrared excess $(W1-W2=0.993, W3-W4=2.675)$ typical for young stellar objects (Koenig and Leisawitz 2014) or quasars (Yan et al. 2013). This object is also present in the catalog of young stellar objects (Marton et al. 2016), where the selection was made based on the WISE and 2MASS colors.
To clarify the nature of the source, we performed optical spectroscopic observations with the AZT-33IK telescope. The exposure time was 50 min. Despite the low flux (r$\sim$23 from the PanSTARRS data), we were able to obtain a significant spectrum of the source. A redshifted broad $H_{\alpha}$ emission line is clearly seen at a wavelength of 7123.2 $\angstrom$ , suggesting that the object being investigated is most likely the active nucleus of a Seyfert 1 galaxy at redshift $z = 0.085\pm0.005 $($D_L\simeq399 Mpc$) (Fig. 1). Apart from the hydrogen line, a narrow N\,II line is seen in the spectrum, which is also typical for objects of this class. The fact that, according to the NVSS (21 cm) data, the source exhibits a radio activity, which is quite typical for active galactic nuclei (AGNs), can serve as an additional confirmation for an extragalactic nature of the object.

\begin{figure*}[t]
\centering
\includegraphics[width=0.82\columnwidth,trim={0cm 0 0cm 0cm},clip]{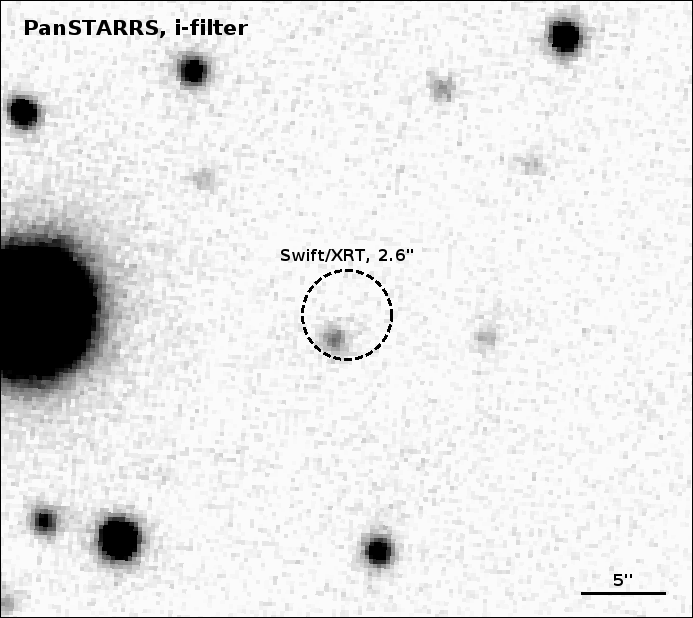}
\includegraphics[width=0.82\columnwidth,trim={0cm 0cm 0cm 0cm},clip]{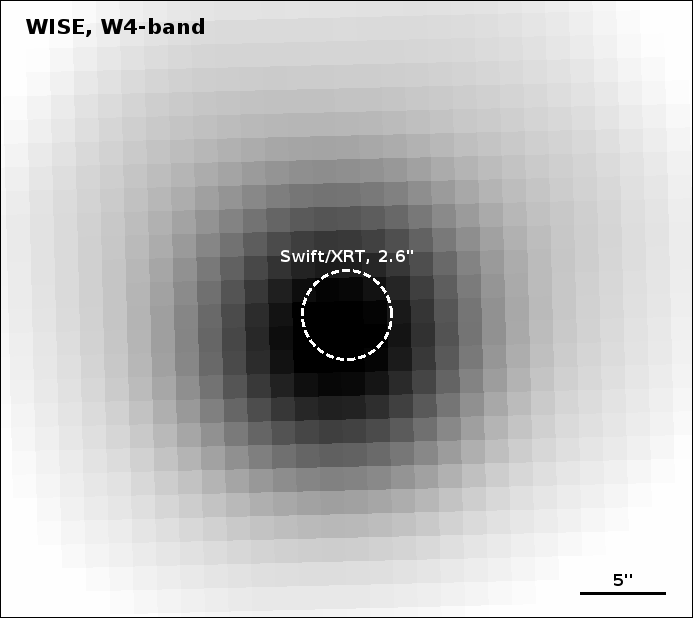}
\includegraphics[width=0.82\columnwidth,trim={1cm 7cm 0cm 2cm},clip]{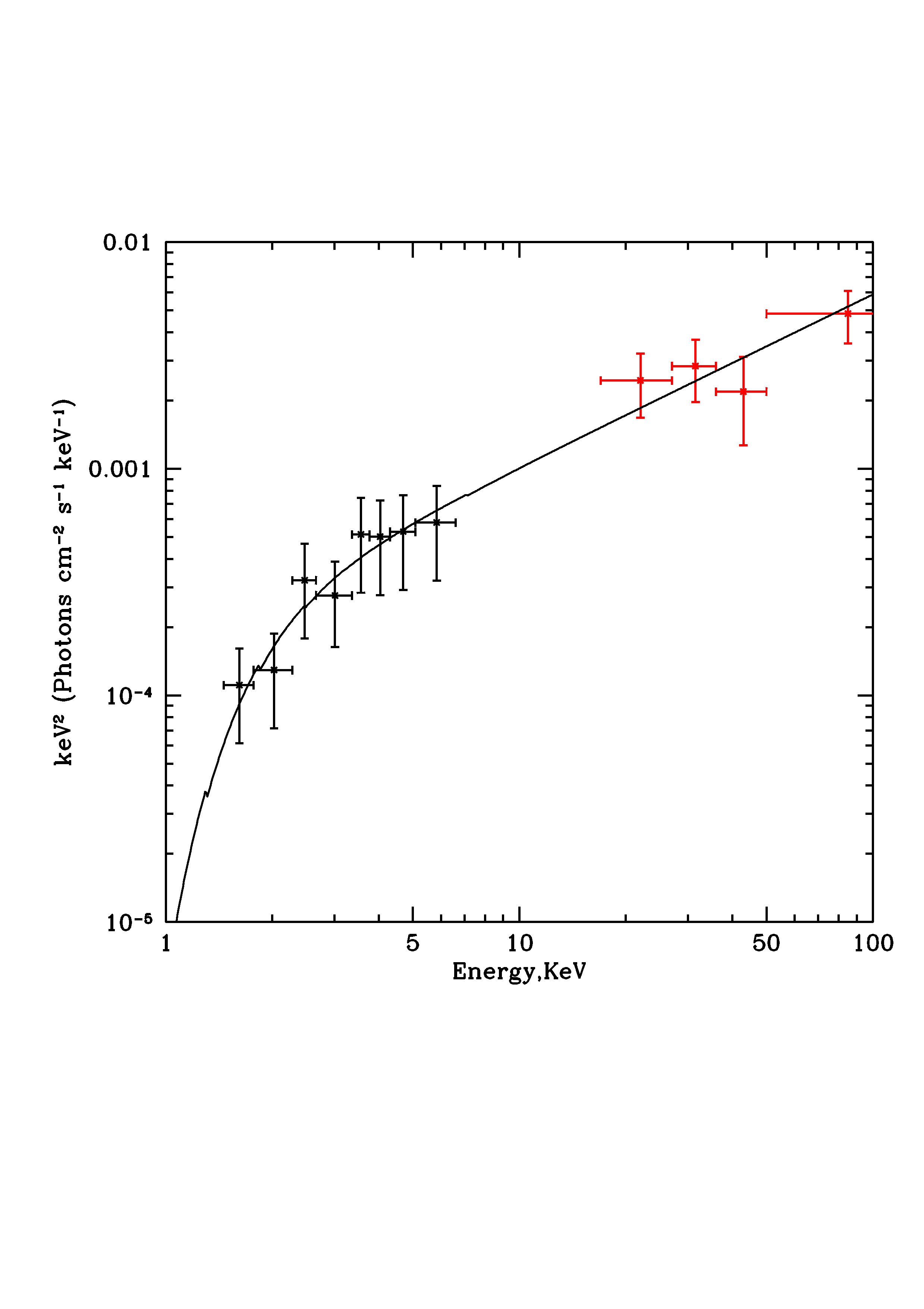}
\includegraphics[width=0.82\columnwidth,trim={0cm 0cm 0cm 0cm},clip]{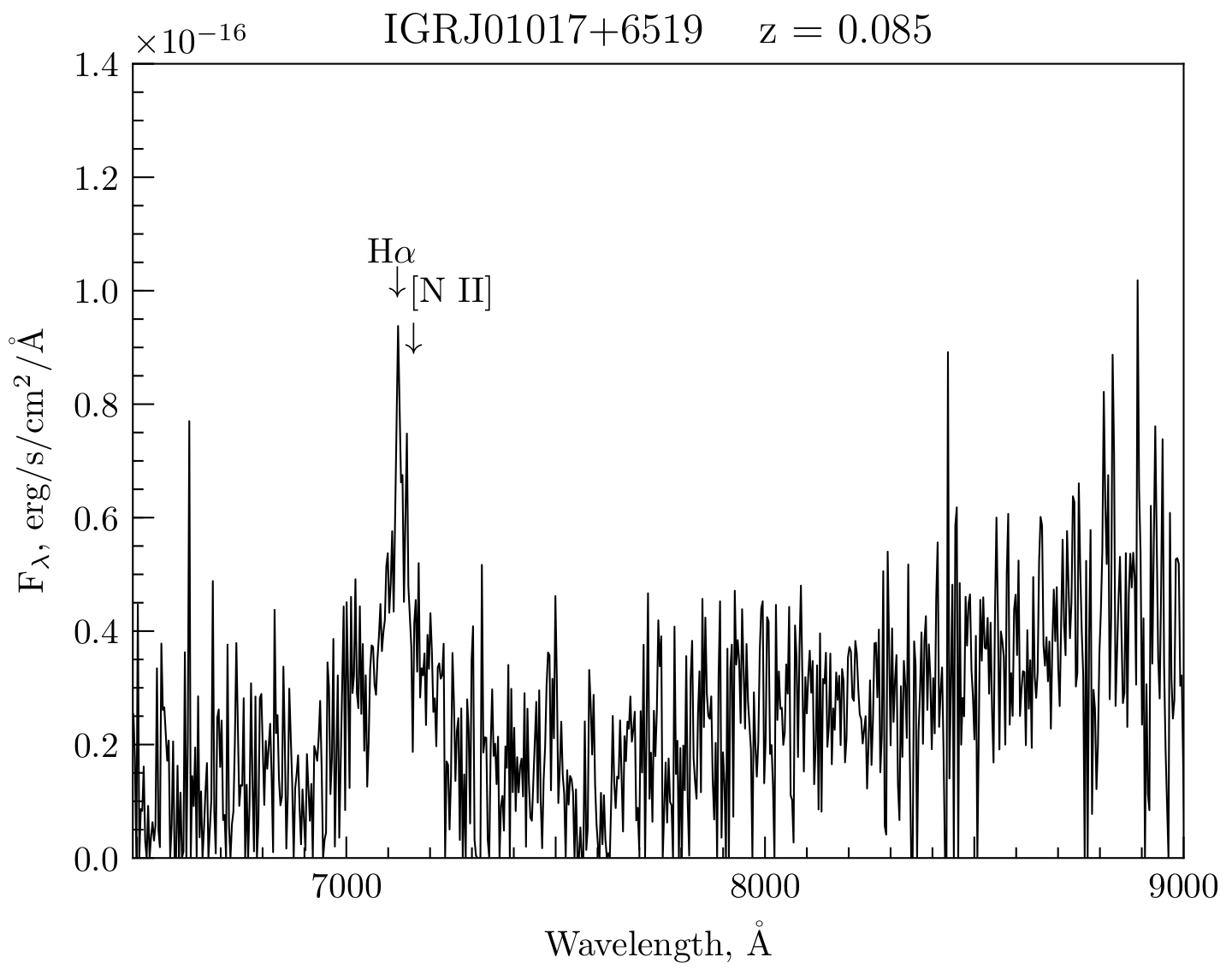}

\caption{(Top) Images of the sky region containing IGR\,J01017+6519 obtained in the PanSTARRS sky survey in the i filter and the ALLWISE sky survey in the W4 band. The dashed circles mark the XRT/Swift localization of the soft X-ray source (Table 1). (Bottom left) The X-ray spectrum of IGR\,J01017+6519 reconstructed from the XRT/Swift (black dots) and IBIS/INTEGRAL (red dots) data. The solid line indicates the fit to the spectrum by a power law with absorption. (Bottom right) The optical spectrum of IGR\,J01017+6519 obtained with the AZT-33IK telescope. The most significant emission lines are marked.}  \label{fig:IGR1}
\end{figure*}

\subsection{IGR\,J08215-1320}

Two galaxies, NGC\,2578 and MCG-02-22-003 (Skrutskie et al. 2006), that were marked in Krivonos et al. (2017) as possible identifications of the object fall into the INTEGRAL position error circle of IGR\,J08215-1320. We requested additional XRT/Swift observations of the source’s fields (ObsID 00010023001, 00010023002, March 2017). According to the results obtained, both these galaxies are detected as faint X-ray sources with comparable fluxes. Thus, each of them could be a counterpart of IGR J08215-1320. However, if we compare the sky images constructed in the 0.5--3 and 6--10 keV energy bands (see Fig. 2), then it becomes clear that the galaxy NGC\,2578 that we see edge-on is a significantly softer X-ray source than the galaxy MCG-02-22-003 that we see face-on. Thus, the most probable counterpart of IGR\,J08215-1320 is the galaxy MCG-02-22-003 (PGC 023449) that, according to the 6dF spectroscopic database (g0821335-132104; Jones et al. 2009), has a redshift $z = 0.0144 (D_{L}\simeq62 Mpc)$. According to this database, it possesses a set of characteristic features, such as narrow Balmer, N\,II, and O\,III emission lines, and is a Seyfert 2 galaxy.

\begin{figure*}[t]
\centering
\includegraphics[width=0.82\columnwidth,trim={0cm 0 0cm 0cm},clip]{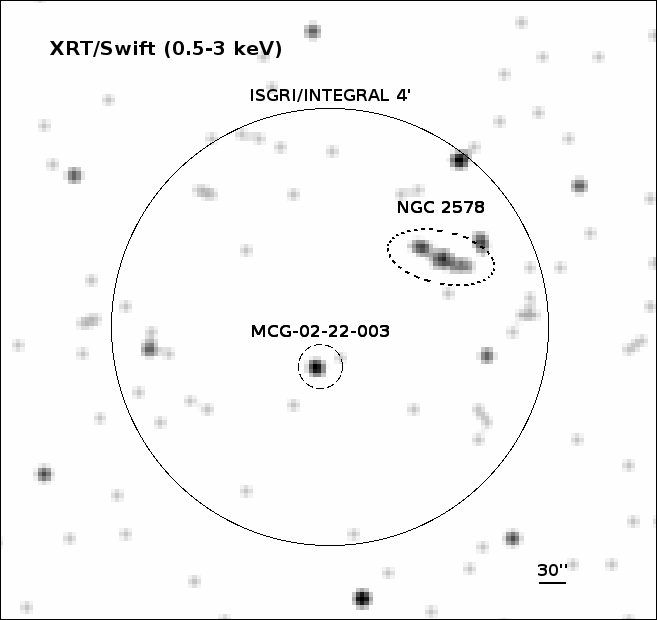}
\includegraphics[width=0.82\columnwidth,trim={0cm 0 0cm 0cm},clip]{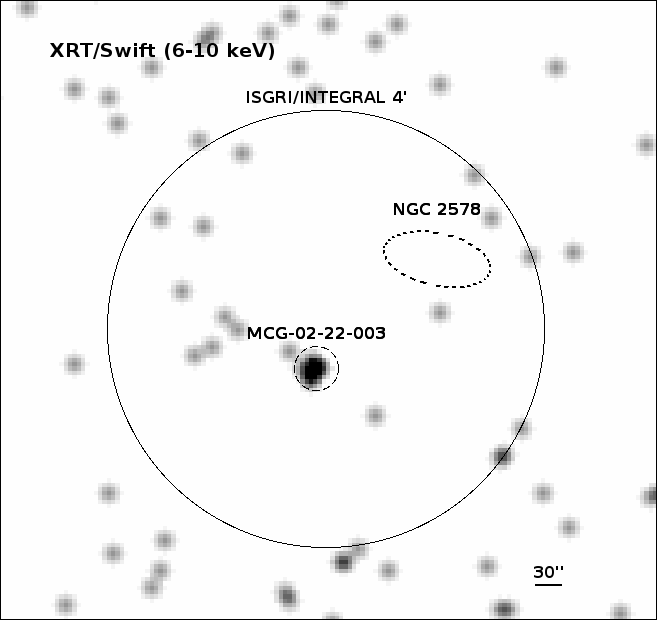}
\includegraphics[width=0.82\columnwidth,trim={0cm 0cm 0cm -1cm},clip]{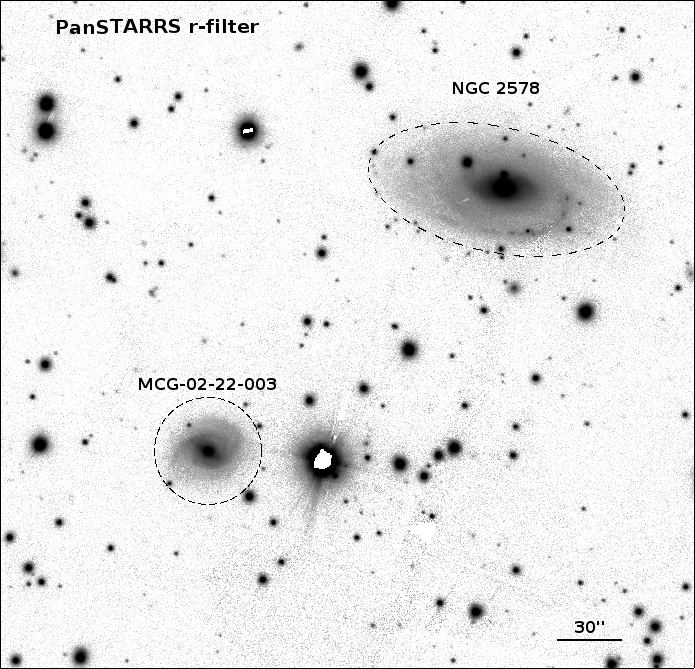}
\caption{(Top) XRT/Swift images of the sky region around IGR\,J08215-1320 in two energy bands. The solid circle indicates the INTEGRAL error circle of the source. (Bottom) PanSTARRS image of the sky region around IGR\,J08215-1320 in the r filter. The dashed circle marks the position and size of the galaxy MCG-02-22-003, while the dashed oval marks the position and size of the galaxy NGC\,2578 inferred from the optical observations.}  \label{fig:IGR2}
\end{figure*}

\subsection{IGR\,J08321-1808}
Using the archival IBIS/INTEGRAL and XRT/Swift observations (ObsID: 00046937001) of this object, we improved its localization and obtained its spectrum in a wide X-ray energy range (Table 2, Fig. 3). Note that the soft X-ray counterpart of the hard X-ray source was observed previously and is present in the catalogs of sources of the ROSAT sky survey (Voges et al. 1999) and the XRT/SWift 1SXPS catalog of soft X-ray sources (Evans et al. 2014).

The high XRT positional accuracy of IGR\,J08321-1808 allowed the source to be unequivocally identified in the optical band using PanSTARRS sky maps. It can be seen from Fig. 3 that only one fairly bright object with $r\simeq16.76$ (PanSTARRS) falls into the error circle. Further observations of this object with RTT-150 allowed its spectrum to be obtained, where narrow [O\,III] 4959, 5007 lines and a bright broad emission line that can be identified with the $H_{\alpha}$ line at $z = 0.135\pm0.001 (D_L\simeq655 Mpc)$ are clearly distinguishable. All these features are typical for Seyfert 1 galaxies.

\begin{figure*}[t]
\centering
\includegraphics[width=0.80\columnwidth,trim={0cm -2cm 0cm 0cm},clip]{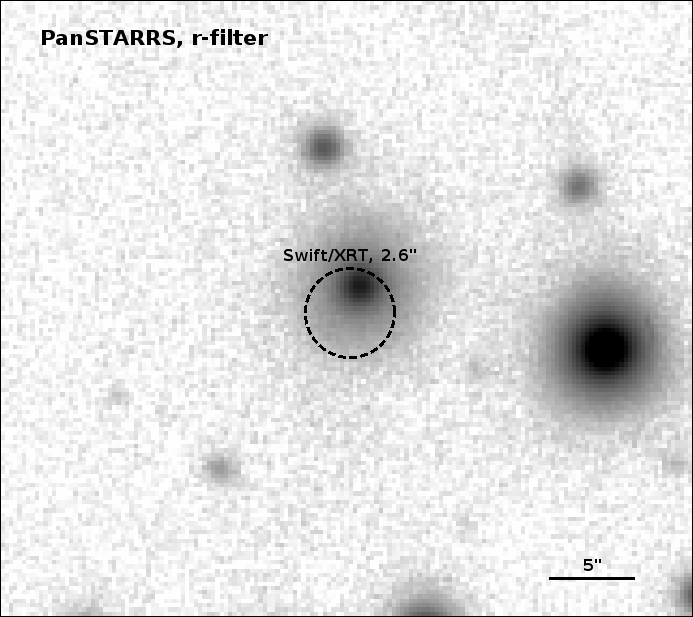}
\includegraphics[width=0.95\columnwidth,trim={0cm 7cm 0cm 5cm},clip]{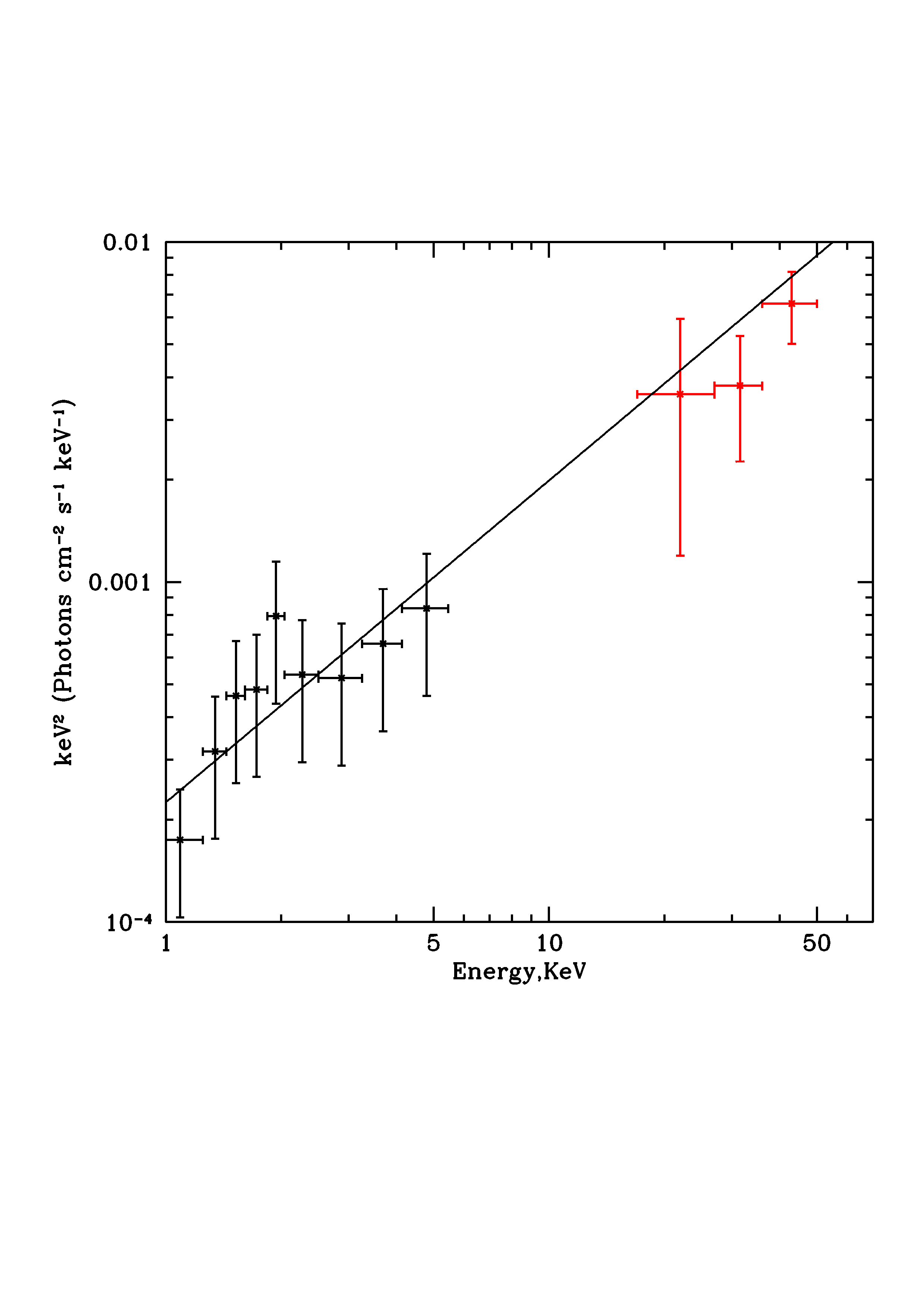}
\includegraphics[width=0.99\columnwidth,trim={0.5cm 1cm 1cm 5cm},clip]{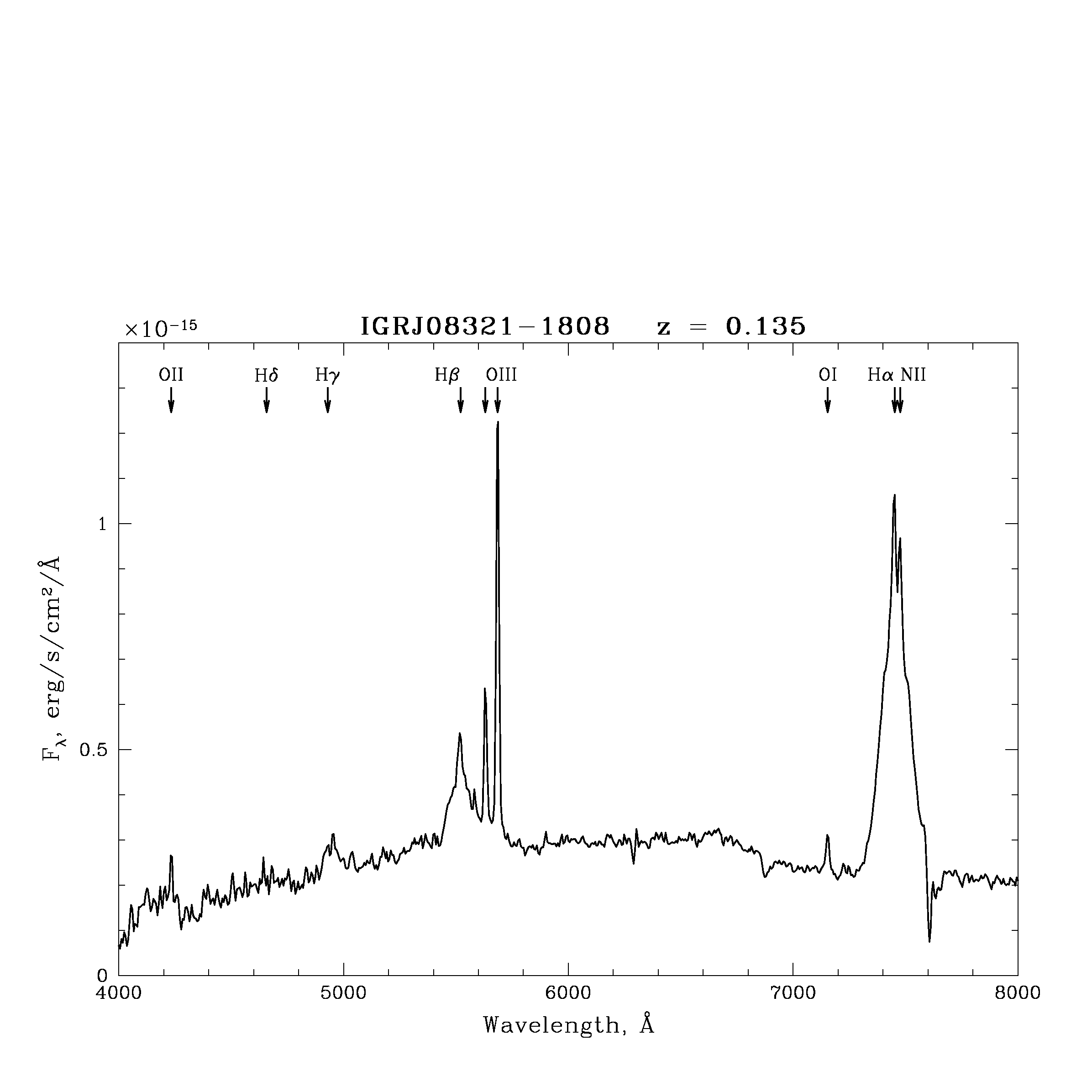}

\caption{(Top left) An image of the sky region containing IGR\,J08321-1808 obtained in the PanSTARRS sky survey in the r-filter. The dashed circle marks the XRT/Swift localization of the soft X-ray source (Table 1). (Top right) The X-ray spectrum of IGR J08321-1808 from the XRT/Swift (black dots) and IBIS/INTEGRAL (red dots) data. The solid line indicates the fit by a simple power law. (Bottom) An optical spectrum of IGR\,J08321-1808. The arrows on the upper and lower left panels mark the most significant emission lines and the most significant emission lines of the optical spectrum, respectively.}  \label{fig:IGR3}
\end{figure*}

\subsection{IGR\,J11299-6557}
To improve the localization of the object, we requested the XRT/Swift observations (ObsID 00010544001, February 2018), based on which we were able to determine the soft X-ray counterpart of the source
and to measure its X-ray flux (Table 1). The combined broadband spectrum of the object from the
XRT/Swift and IBIS/INTEGRAL data is well described by a simple power law with absorption (Table 2, Fig. 4).

Owing to the high XRT localization accuracy of the source, we were able to unequivocally identify the infrared counterpart of the object WISE\,J112956.44-655521.8 using WISE data (see Fig. 4 and Table 2), which is a very bright object in the mid-infrared with magnitudes $W2 = 9.969\pm0.020$ and $W4 = 4.475\pm0.021$. To classify the object, we used the selection criterion from Stern et al. (2012) $(W1-W2 > 0.8$ and $W2 < 15.05$) by first correcting the $(W1-W2)$ color of the counterpart for reddening. The reddening correction $E(W1~-~W2)=0.03$ for the sky region toward IGR\,J11299-6557 was taken from the maps in Schlafly and Finkbeiner (2011).

Thus, having obtained the dereddened color of the counterpart $(W1-W2)_0=1.15$, we can provisionally attribute the source to the class of AGNs. It is worth noting that the same conclusion about the nature of WISE\,J112956.44-655521.8 was drawn by Kouzuma and Yamaoka (2010), Edelson and Malkan (2012), and Secrest et al. (2015).

\begin{figure*}[t]
\centering
\includegraphics[width=0.85\columnwidth,trim={0cm 0 -1cm 0cm},clip]{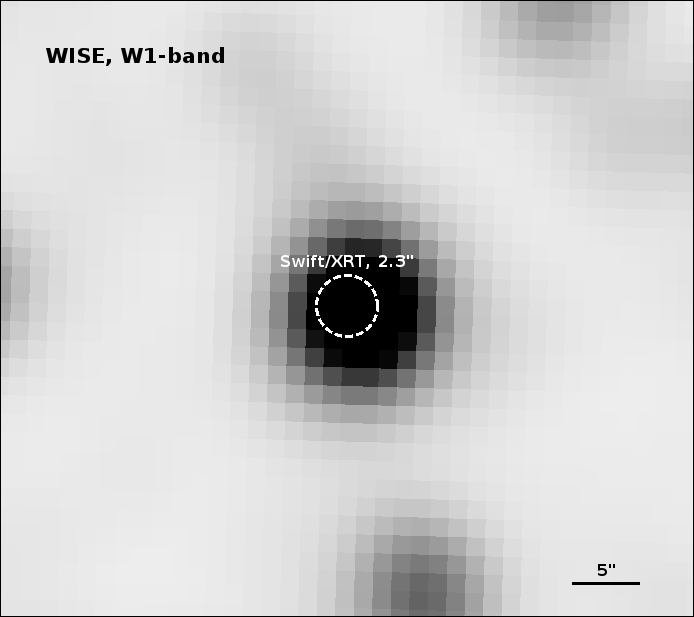}
\includegraphics[width=0.85\columnwidth,trim={-1cm 0 0cm 0cm},clip]{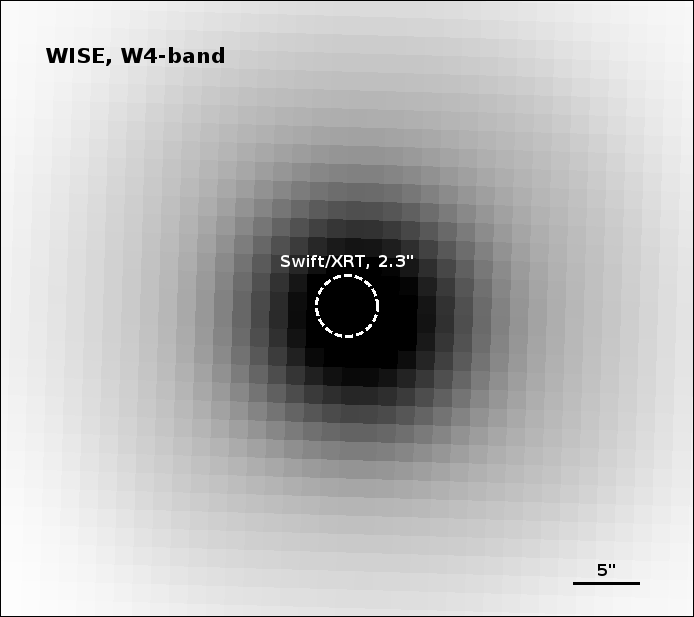}
\includegraphics[width=0.85\columnwidth,trim={0cm 0cm 0cm -2cm},clip]{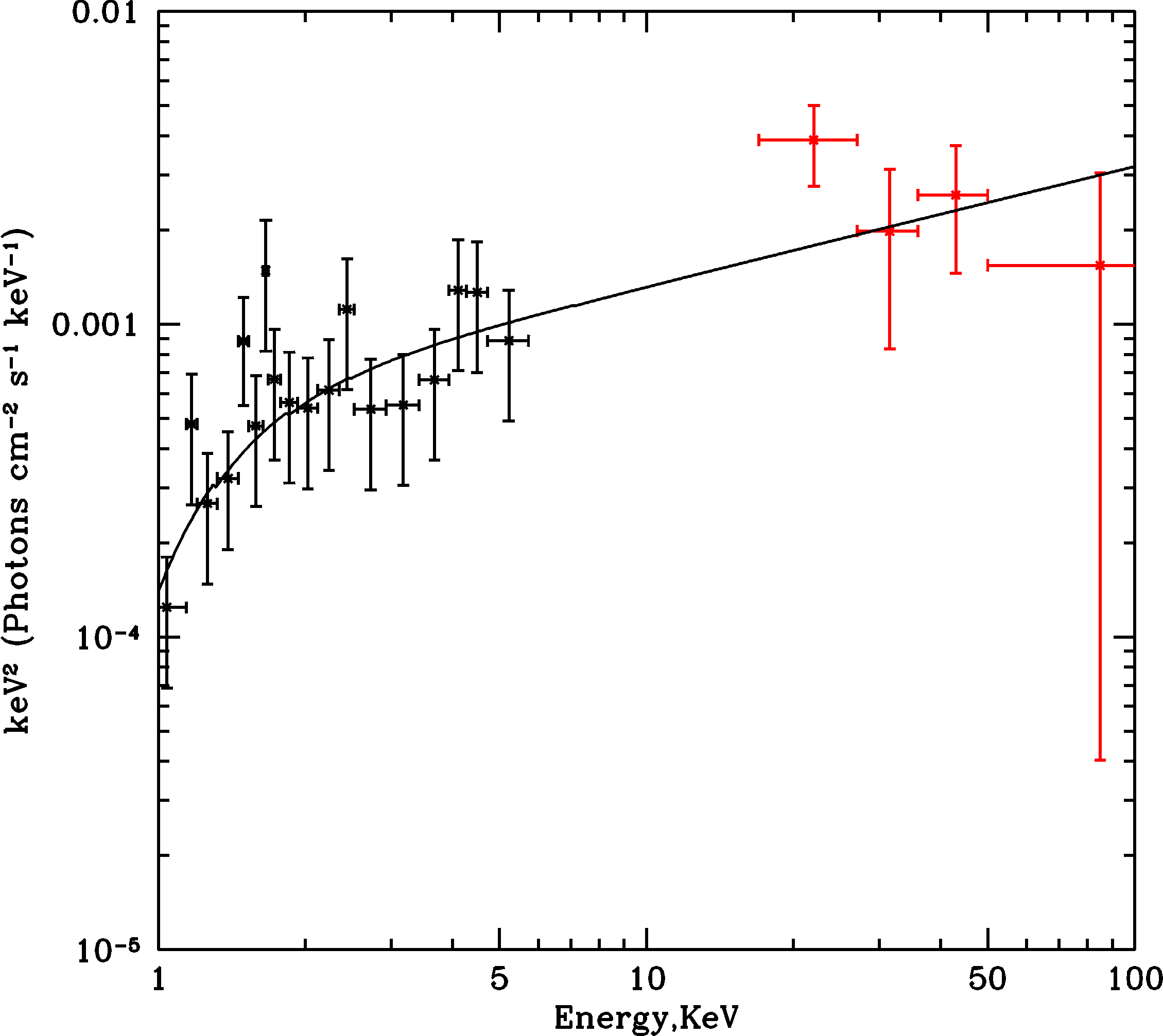}
\caption{(Top) WISE images of the sky region containing IGR J11299-6557 in the W1 and W4 bands. The dashed circles mark the XRT/SWift localization of the soft X-ray source (Table 1). (Bottom) The X-ray spectrum of IGR\,J11299-6557 from the XRT/Swift and IBIS/INTEGRAL data. The solid line indicates the fit to the broadband X-ray spectrum by a simple power law with absorption.}  \label{fig:IGRn1}
\end{figure*}


\subsection{IGR\,J14417-5533}
Owing to the XRT/Swift observations (ObsID 00010546001, February 2018), we were able to improve the localization of this source, which, in turn, allowed the counterpart of the object in the mid-infrared to be determined (WISE\,J144118.74-553335.1, Fig. 5). The XRT/Swift observations also allowed us to measure the flux from the source (Table 1) and, in combination with the IBIS/INTEGRAL data, to construct its broadband spectrum. The constructed spectrum is best described by a power law with absorption (Table 2, Fig. 5).

The object was classified based on the infrared colors for this source using the same algorithm as that for IGR\,J11299-6557. Having derived the dereddened color $(W1-W2)_0=0.95$, we used the criterion from Stern et al. (2012) and obtained evidence that the object being investigated is probably an AGN (see also Edelson and Malkan (2012) and Secrest et al. (2015), where the same conclusion regarding WISE\,J144118.74-553335.1 was drawn).

\begin{figure*}[t]
\centering
\includegraphics[width=0.85\columnwidth,trim={0cm 0 -1cm 0cm},clip]{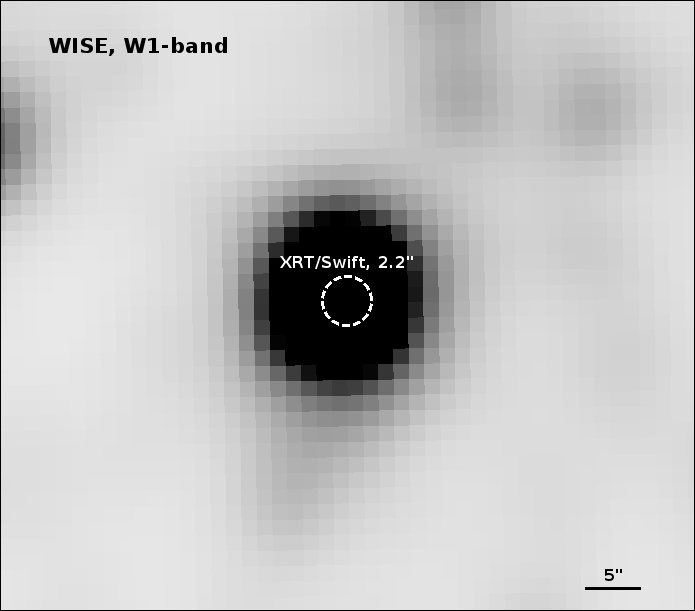}
\includegraphics[width=0.85\columnwidth,trim={-1cm 0 0cm 0cm},clip]{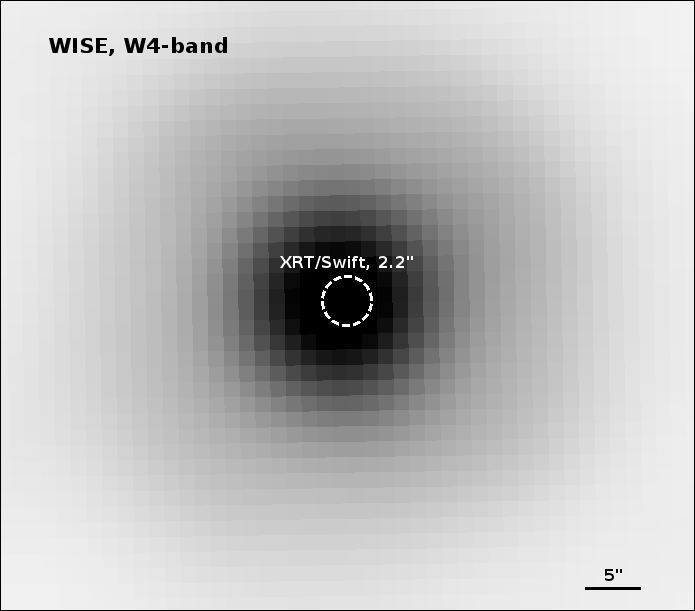}
\includegraphics[width=0.90\columnwidth,trim={1cm 7cm 0cm 3cm},clip]{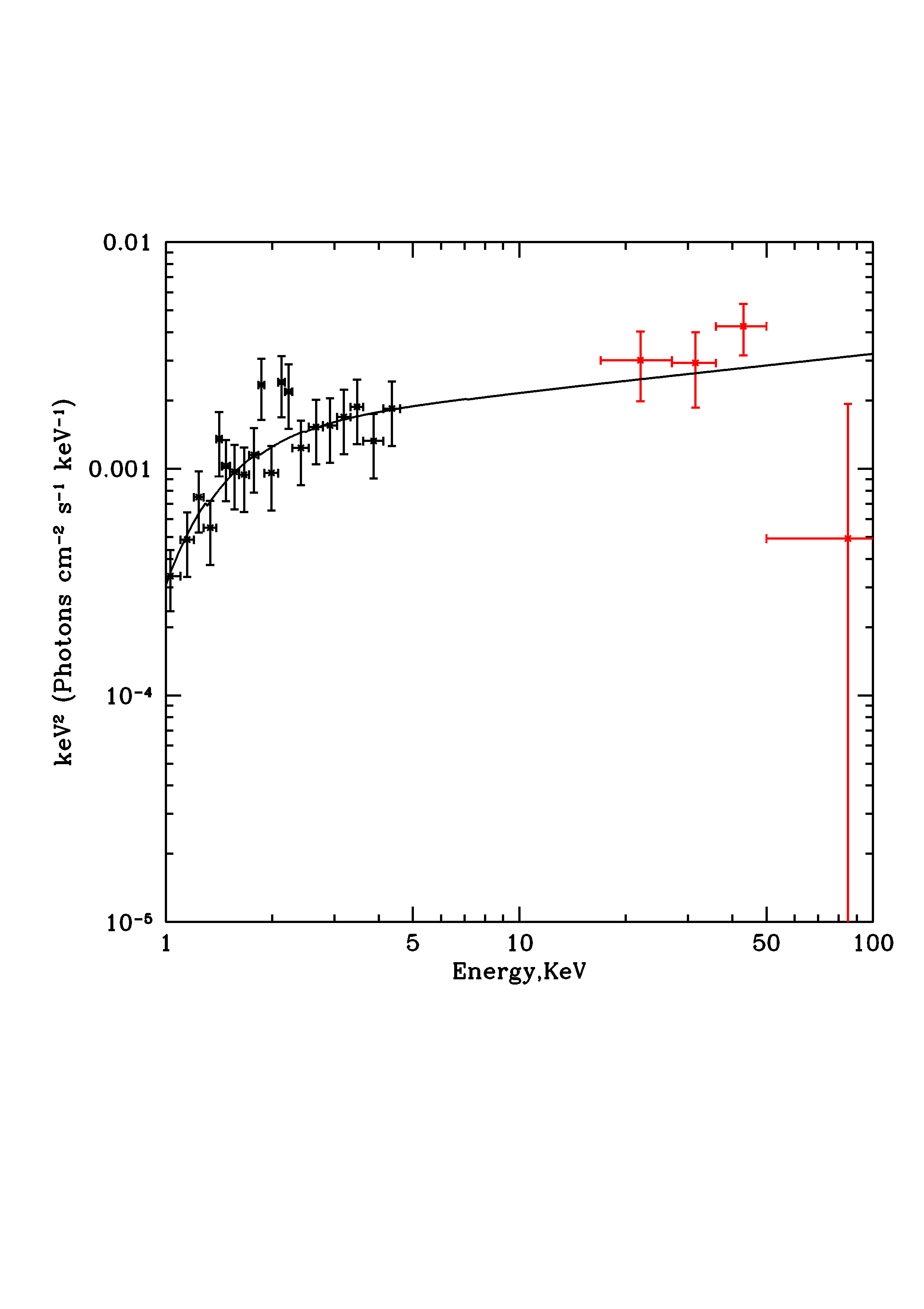}
\caption{Same as Fig. 4 for IGR\,J14417-5533}  \label{fig:IGRn2}
\end{figure*}

\subsection{IGR\,J16494-1740}
Using the publicly accessible XRT/Swift observations of the fields around this source (ObsID 00040989001), we estimated its X-ray flux in the 2–10 keV energy band (see Table 1). The broadband X-ray spectrum of the source is best fitted by a power law with photoabsorption (Table 2, Fig. 6). Owing to the improvement in the position accuracy, we were abale to identify the X-ray source with the galaxy ESO\,586-4 (also known as IRAS\,16464-1733). Note that this object is present under the name 1SXPS\,J164920.9-173840 in the XRT/Swift catalog of soft X-ray sources (Evans et al. 2014). Our estimates of the X-ray flux agree well with its estimates from the above paper. At the same time, the localization accuracy of the object in our case turns out to be higher (probably due to the use of a more up-to-date version of the software). In the case under consideration, this is not so important, because the choice of an optical counterpart is obvious and unambiguous.

Our spectroscopic observations of this galaxy with RTT-150 allowed us to obtain significant spectra of the object (Fig. 6) and to determine its class and redshift. In particular, a narrow $H_{\alpha}$ line and N\,II lines are clearly seen in the spectrum. Hence, we are dealing with a Seyfert 2 galaxy with an active nucleus at red-shift $z = 0.023\pm0.001 (D_L\simeq121 Mpc)$. Our redshift estimates agree well with those from Hasegawa et al. (2000). Note that in Rojas et al. (2017), where the Swift/BAT catalog of sources with identifications is presented, this source is known under the name 2PBC\,J1649.3-1739 and it was also classified as a Seyfert 2 galaxy. Note that such an interesting effect as the absence of oxygen lines in the spectrum of a Seyfert 2 galaxy can be explained by significant additional absorption in the galaxy itself, because we see it edge-on. For this reason, such objects, even if they are located nearby, are fairly difficult to investigate in the optical band. At the same time, they are perfectly suitable for research by X-ray observatories, for example, the INTEGRAL observatory. A number of objects like IGR\,J16494-1740 were previously considered by Burenin et al. (2008).

\begin{figure*}[t]
\centering
\includegraphics[width=0.84\columnwidth,trim={0cm -1cm -1cm 0cm},clip]{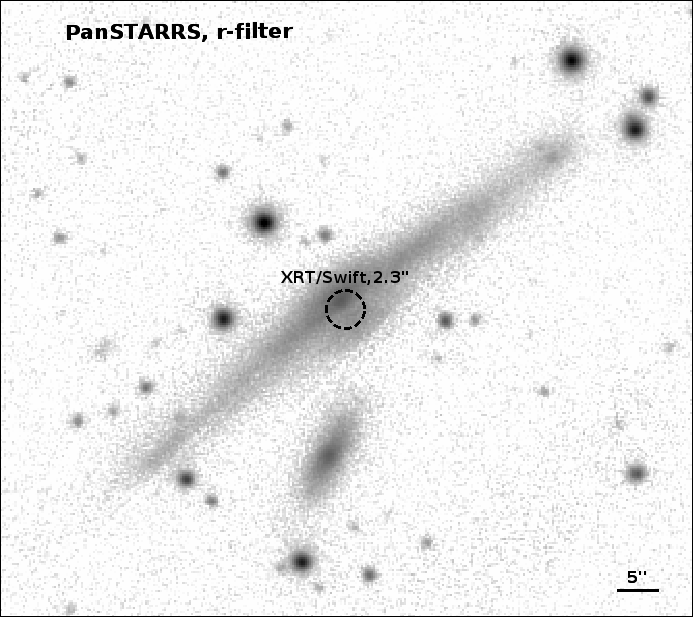}
\includegraphics[width=0.90\columnwidth,trim={-1cm 0cm 0cm 0cm},clip]{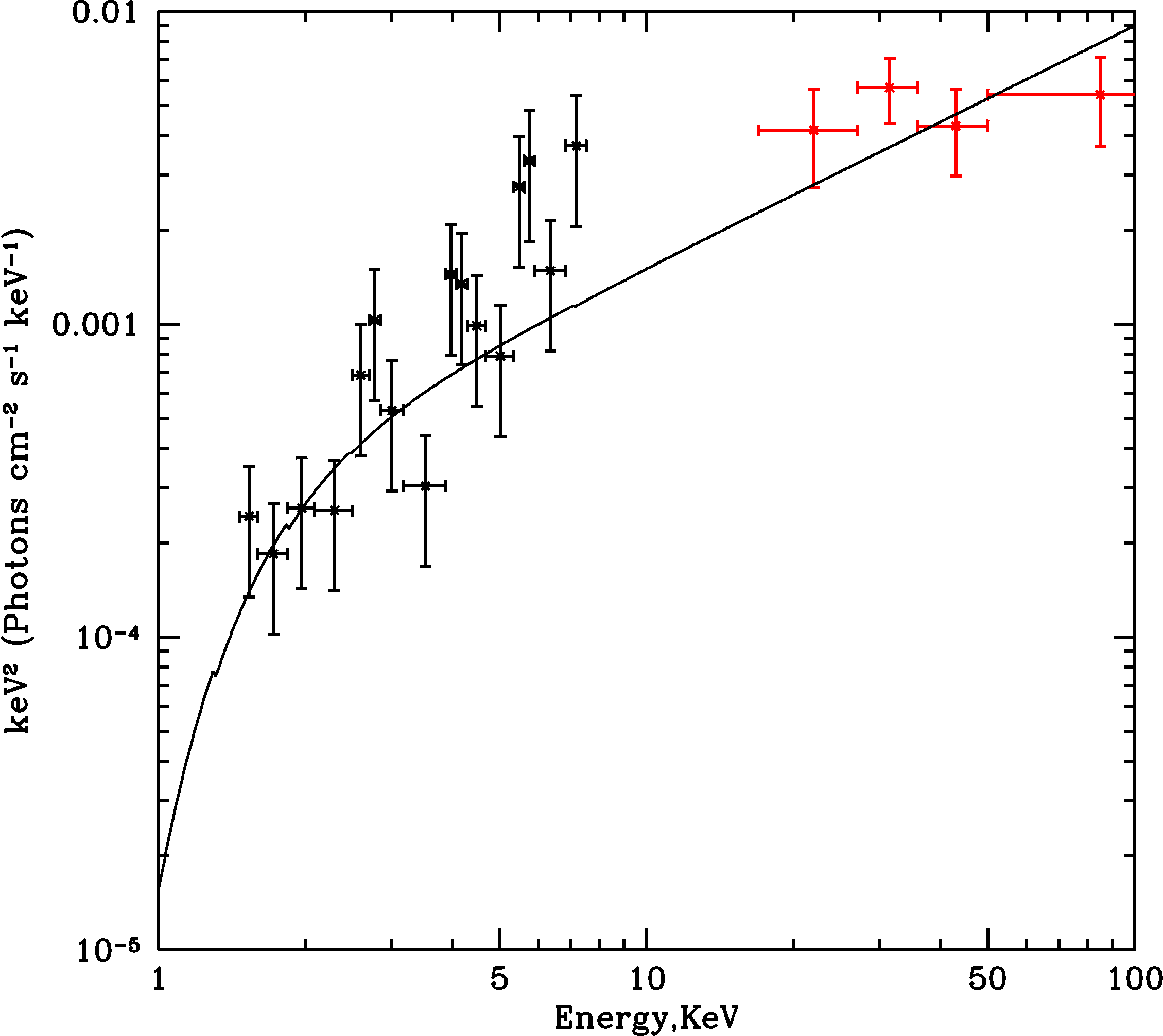}
\includegraphics[width=0.99\columnwidth,trim={1cm 1cm 1cm 4cm},clip]{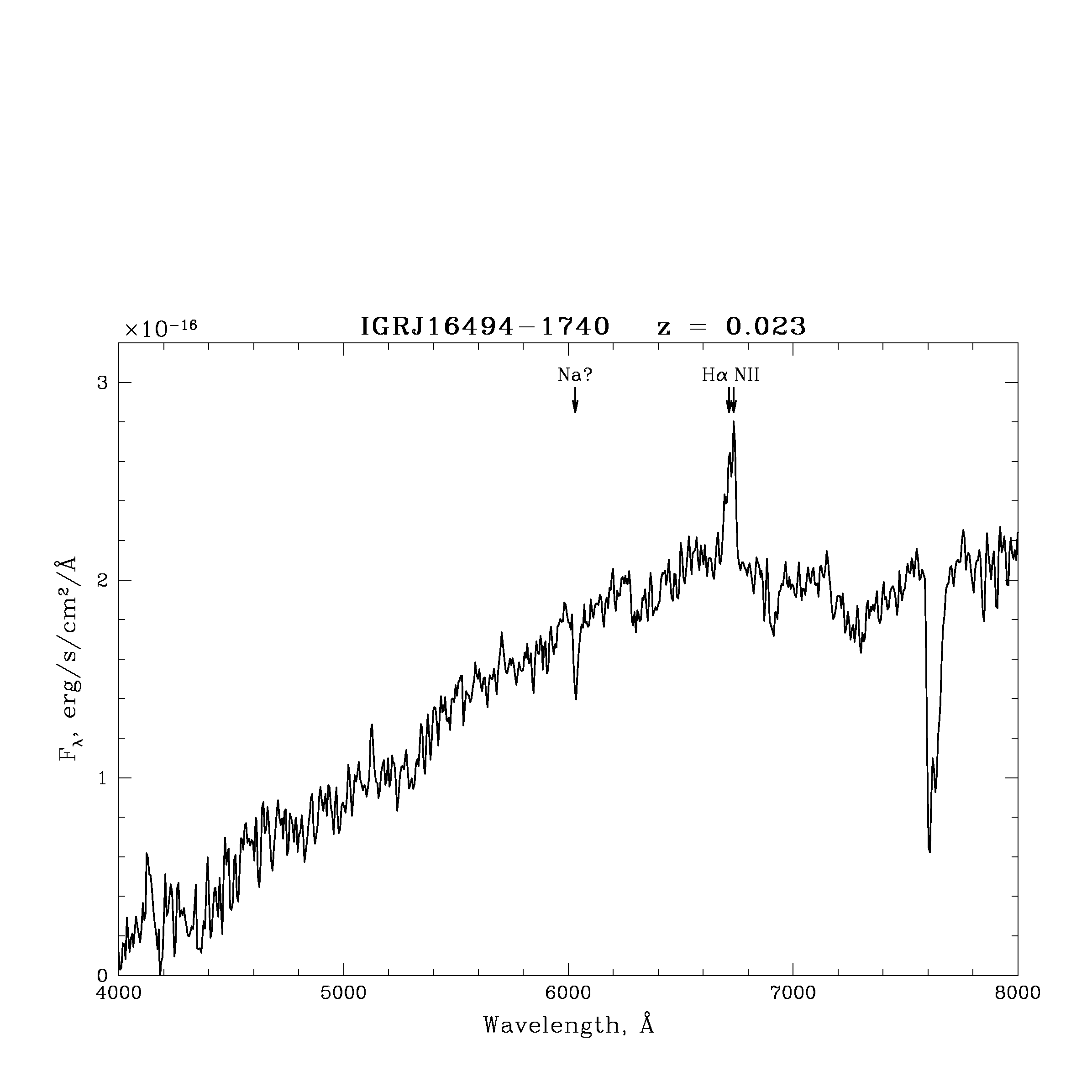}
\caption{Same as Fig. 4 for IGR\,J16494-1740. The black curve on the upper right panel indicates the fit to the broadband X-ray spectrum by a simple power law with absorption.}  \label{fig:IGR4}
\end{figure*}

\subsection{IGR\,J17098-2344}
Analysis of the archival observations of IGR\,J17098-2344 by the INTEGRAL and XMM-Newton space observatories (ObsID 0206990701) allows us to study its spectral properties in a wide X-ray range and to determine its possible optical counterpart (Tables 2 and 3, Fig. 7).
Note that the soft X-ray counterpart of the source was detected previously by the ROSAT observatory(Voges et al. 1999) and is known under the name 1RXS\,J170944.9-234658.
Using our spectroscopic observations with RTT-150, we were able to determine the redshift of the optical counterpart $z = 0.036\pm0.001 (D_L\pm163 Mpc)$, which agrees well with the estimates of $z\pm 0.0364$ from Durret et al. (2015) devoted to the spectroscopic studies of a large sample of galaxies. The study of spectral lines allowed the true class of the source to be determined. On the one hand, in contrast to the previous objects, here we ran into some ambiguity in interpreting the spectrum obtained. On the other hand, according to our observations, the line flux ratio $F_{H\beta} /F_{O\,III}$ for IGR\,J17098-2344 is about 3.7, which defines this object as an Sy1.2 galaxy. On the other hand, the equivalent width of the $H{\beta}$ line is less than 1200 km s$^{-1}$ and, in addition, a significant emission in Fe\,II lines is detected in the spectrum of IGR\,J17098-2344. The latter is typical for narrow-line Seyfert 1 galaxies, NLSy1 (Osterbrock and Pogge 1985; Goodrich 1989).
Note that the object 2PBC\,J1709.7-2349 with coordinates close to those of IGR\,J17098-2344 is present in the catalog of hard X-ray sources detected
by the BAT/Swift telescope. Its optical identification made by Rojas et al. (2017) coincides with our results for IGR\,J17098-2344, suggesting that both these sources are the same object. These authors classified the counterpart of the source as Sy1.2.

Note also that for the best description of the source’s broadband X-ray spectrum we had to add an exponential cutoff at a relatively low energy to the model of a power law with absorption, which is not characteristic for typical AGNs. Furthermore, the additional factor that allows the normalizations of the two data sets to be equalized differs significantly from unity and is $0.53\pm0.16$, which may suggest the object’s variability.

Thus, it can be asserted with confidence that the soft X-ray source detected within the INTEGRAL error circle is an AGN. However, it is impossible to unambiguously assert that precisely this source is a soft X-ray counterpart of IGR\,J17098-2344. This object requires a separate additional study, for example, by the Nustar observatory.

\begin{figure*}[t]
\centering
\includegraphics[width=0.79\columnwidth,trim={0cm -1.5cm 0cm 0cm},clip]{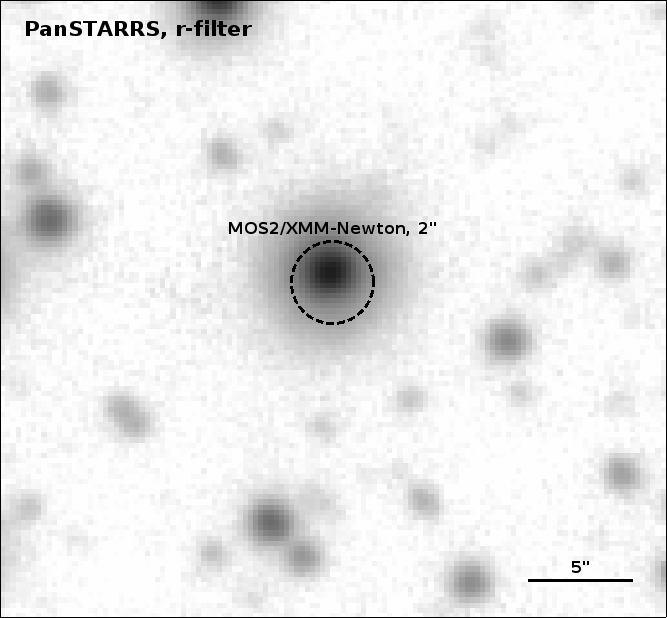}
\includegraphics[width=0.97\columnwidth,trim={0cm 7cm 0cm 2cm},clip]{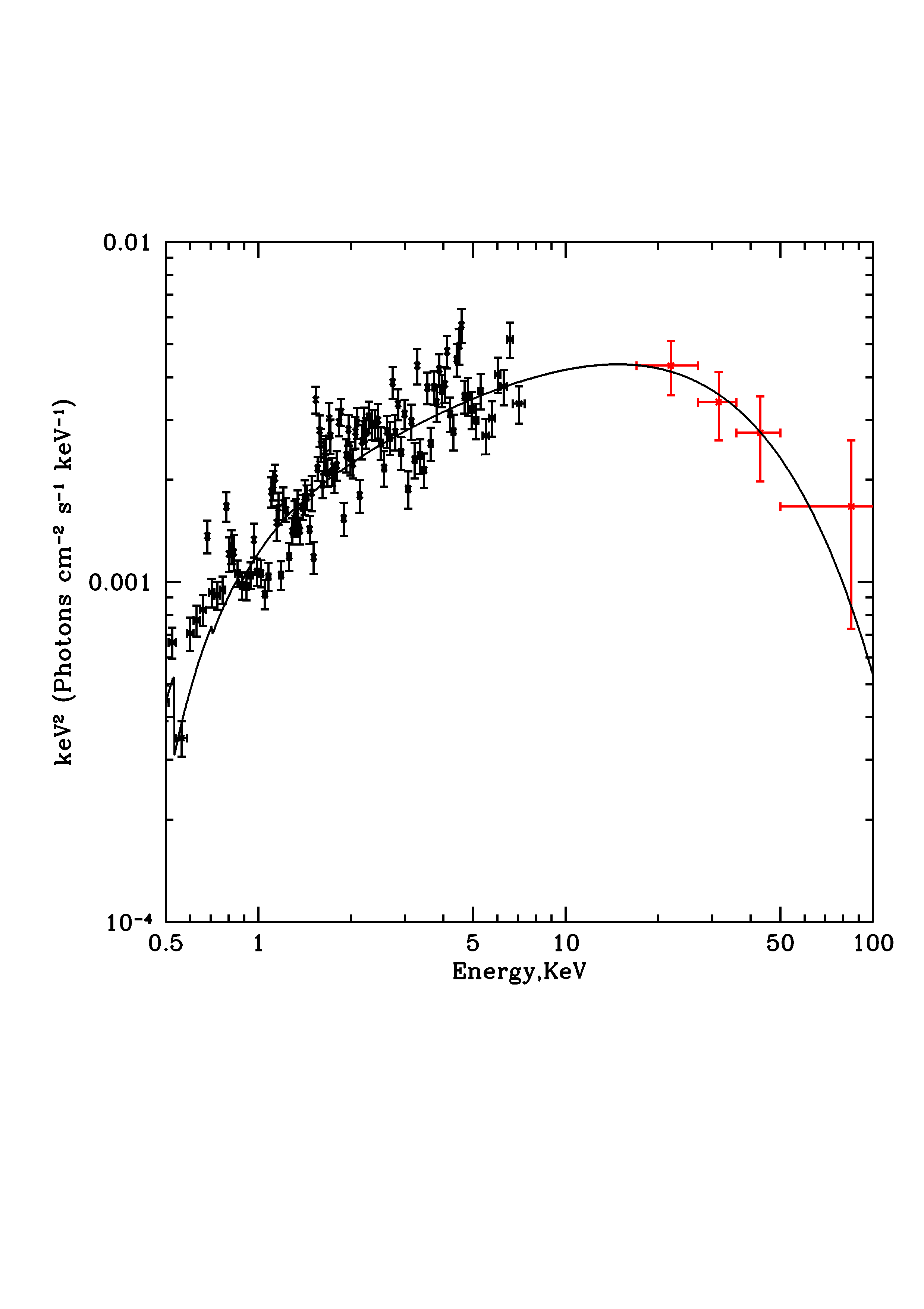}
\includegraphics[width=0.93\columnwidth,trim={1cm 1cm 1cm 4cm},clip]{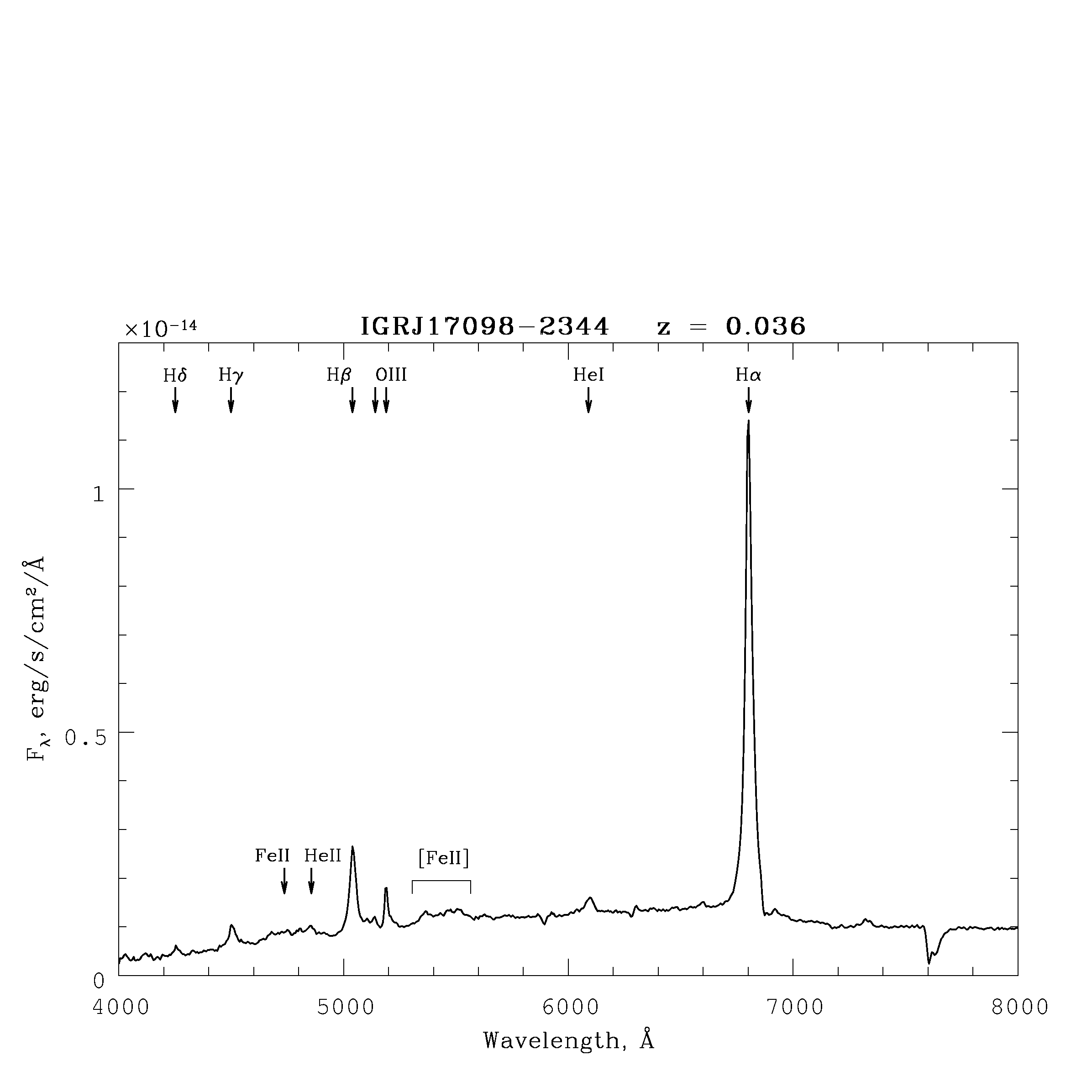}
\caption{Same as Fig. 4 for IGR\,J17098-2344. The X-ray spectrum of the source was obtained from the MOS2/XMM-Newton (black dots) and IBIS/INTEGRAL (red dots) data. The solid curve is the fit to the X-ray spectrum by a power law with an exponential cutoff.}  \label{fig:IGR5}
\end{figure*}
\begin{figure*}[!t]
\centering
\includegraphics[width=0.79\columnwidth,trim={0cm -1.5cm 0cm 0cm},clip]{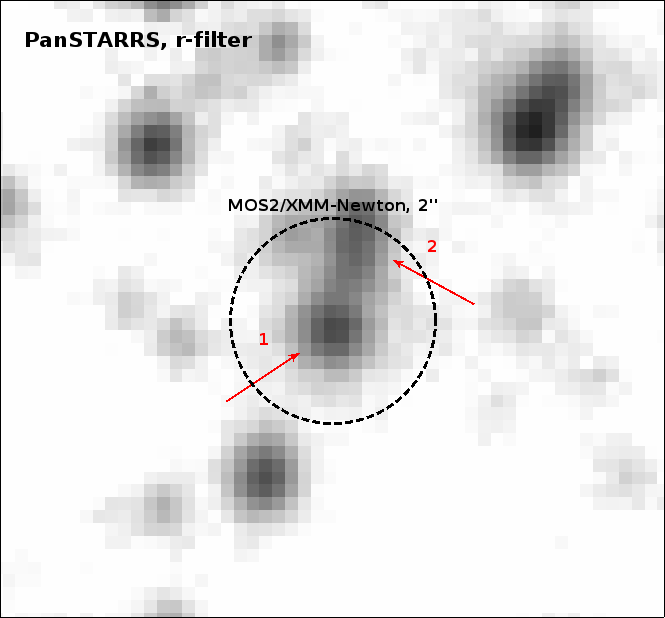}
\includegraphics[width=0.97\columnwidth,trim={0cm 7cm 0cm 2cm},clip]{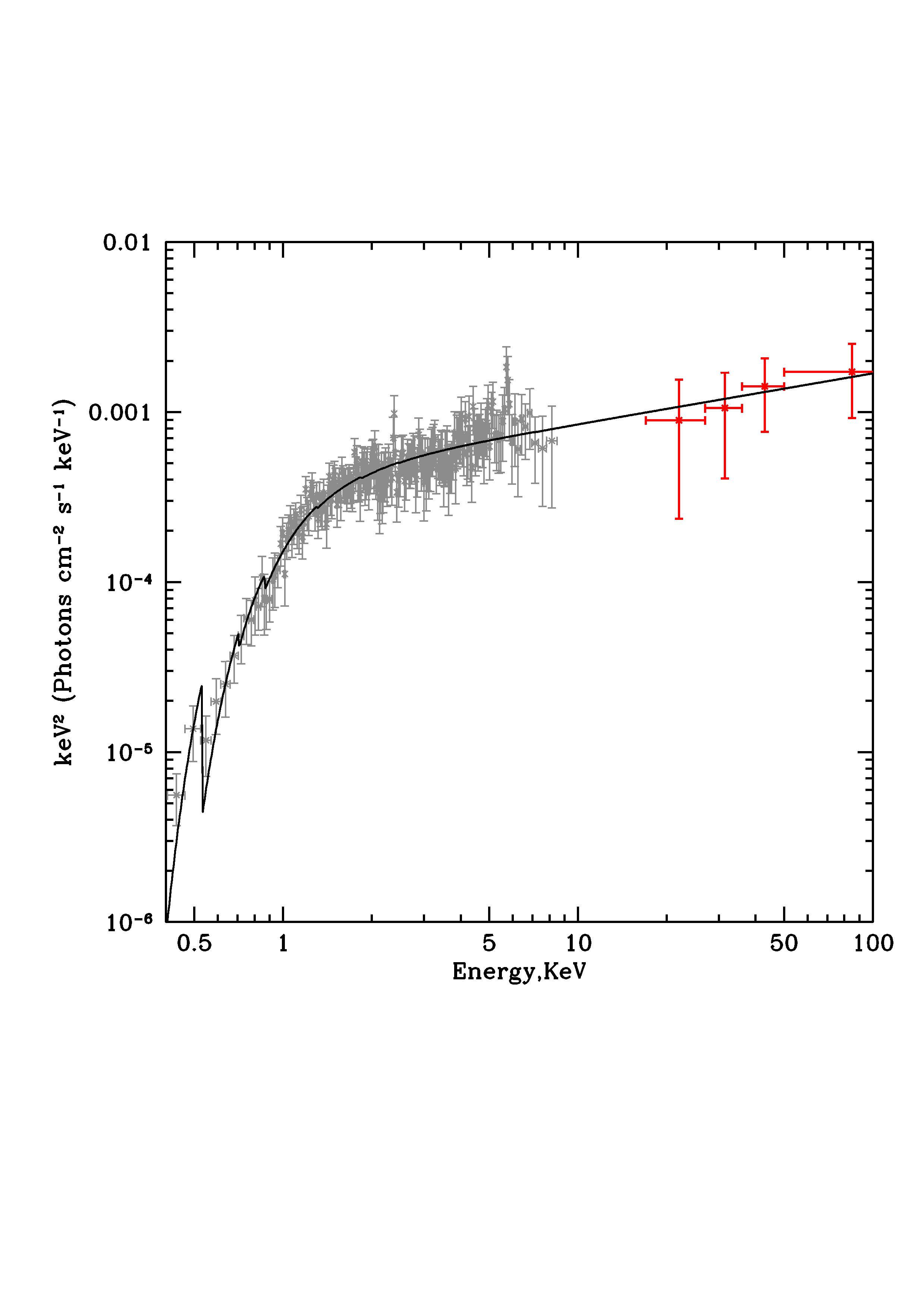}
\includegraphics[width=0.93\columnwidth,trim={1cm 1cm 1cm 4cm},clip]{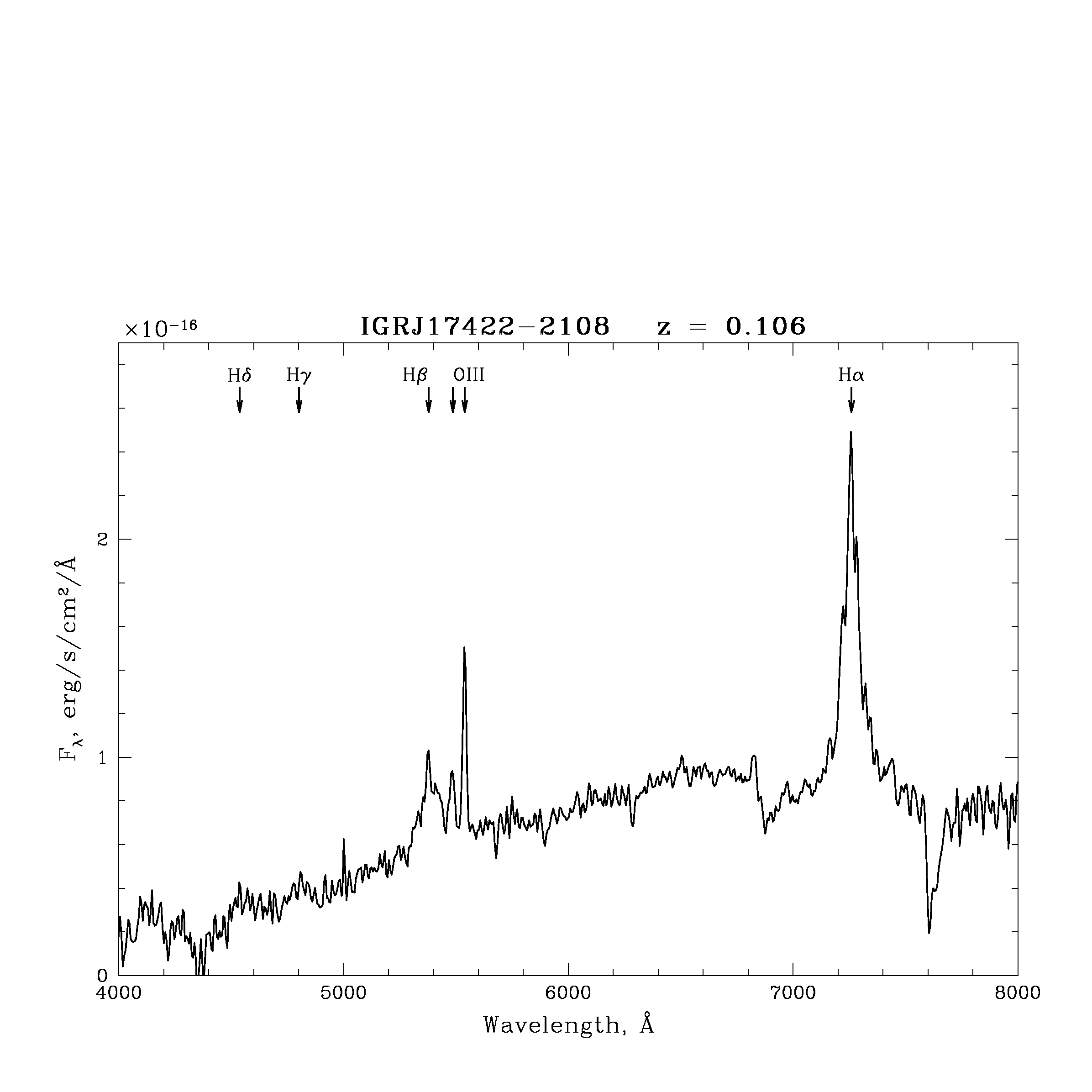}
\caption{Same as Fig. 4 for IGR\,J17422-2108. The solid line is the fit to the X-ray spectrum by a simple power law with absorption. The arrows mark the source’s possible optical counterparts.}  \label{fig:IGR6}
\end{figure*}

\subsection{IGR\,J17422-2108}
Owing to the fact that the sky region containing
IGR\,J17422-2108 fell into the XMM-Newton field of view during the session of observations of the pulsar PSR\,J1741-2054 (ObsID 0693870101), we were able to improve the localization of this object and, having combined these data with the INTEGRAL data, to construct its broadband spectrum (Table 2, Fig. 8). It is clear from the XMM-Newton observations that only one soft X-ray source, which we will consider as a counterpart of IGR\,J17422-2108, falls into the INTEGRAL error circle. Despite the fact that the source is at the edge of the MOS2/XMM-Newton field of view, it is bright enough for its position to be determined with a $\sim2\arcsec$ accuracy.

The X-ray spectrum of this object reconstructed from the above data is well described by a power law with a slope $\Gamma = 1.70\pm0.05$ supplemented by photoelectron absorption. Note that such a slope is typical for extragalactic objects. The introduction of an additional factor equal to $1.49\pm0.44$ in the model was required to achieve the best agreement between the data obtained by different observatories, which may suggest that this object is variable.

Simultaneously two relatively faint optical objects fall into the XMM-Newton position error circle of the source (see Fig. 8). Using RTT-150, we obtained a spectrum for the optical object closer to the center of the XMM-Newton error circle of IGR\,J17422-2108. A broad redshifted $H_{\alpha}$ emission line and a number of other features, including narrow [O\,III] 4959, 5007 lines, are clearly seen in this spectrum. They confirm an extragalactic nature of the object and allow it to be classified as a Seyfert 1 galaxy at redshift $z = 0.106\pm0.001 (D_L\simeq505 Mpc)$. Thus, we can compare the soft X-ray source detected by XMM-Newton with this optical object and classify IGR\,J17422-2108 as an AGN.

Note that the soft X-ray counterpart of IGR\,J17422-2108 investigated in this section is also known under the name 1SXPS\,J174211.7-210354 (Evans et al. 2014). However, according to the data from this catalog, the source’s position is shifted by $5\arcsec$ relative to the position of the source detected by XMM-Newton. This, along with the more significant localization uncertainty (3.7$\arcsec$), did not allow a reliable optical identification of this object to be made previously in such a crowded part of the sky.

\subsection{IGR\,J18044-1829}
Owing to the XRT/Swift observations (ObsID 00010169001,00010169002; June 2017), we were able to improve the localization of this source, to measure its X-ray flux, and to construct its X-ray spectrum. The combined broadband spectrum of the source can be fitted by a power law with an exponential cutoff (Table 2, Fig. 9), which is typical for Galactic objects. Note that IGR\,J18044-1829 is close to the Galactic plane ($l\approx11.19^{\circ}, b\approx1.53^{\circ}$) in a fairly crowded field, and, therefore, several star-like objects, each of which could be an optical counterpart of the source, fall into the XRT error circle. The sources detected at a statistically significant level in the mid-infrared do not fall into the XRT/Swift error circle, which also complicates the determination of the true counterpart of IGR\,J18044-1829. Thus, further studies by X-ray telescopes with a better positional sensitivity, for example, by the HRC/Chandra telescope, are needed for an unequivocal identification of the object being investigated.
\begin{figure*}[t]
\centering
\includegraphics[width=0.85\columnwidth,trim={0cm 0 -1cm 0cm},clip]{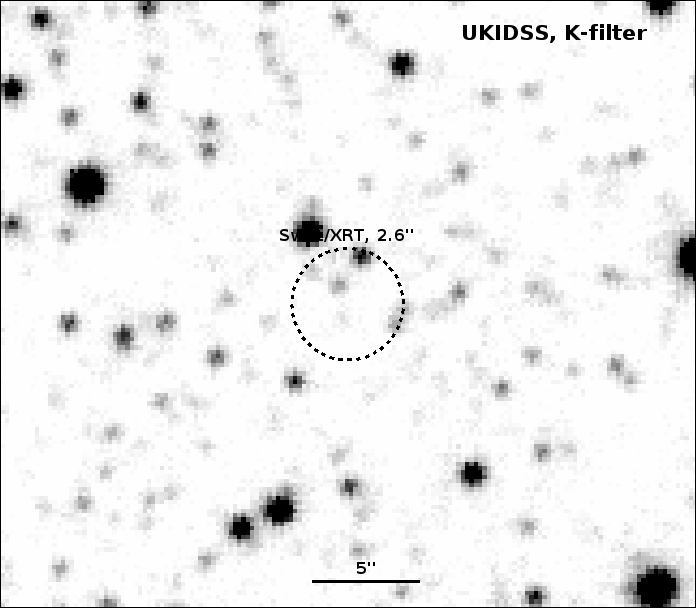}
\includegraphics[width=0.85\columnwidth,trim={-1cm 0cm 0cm 0cm},clip]{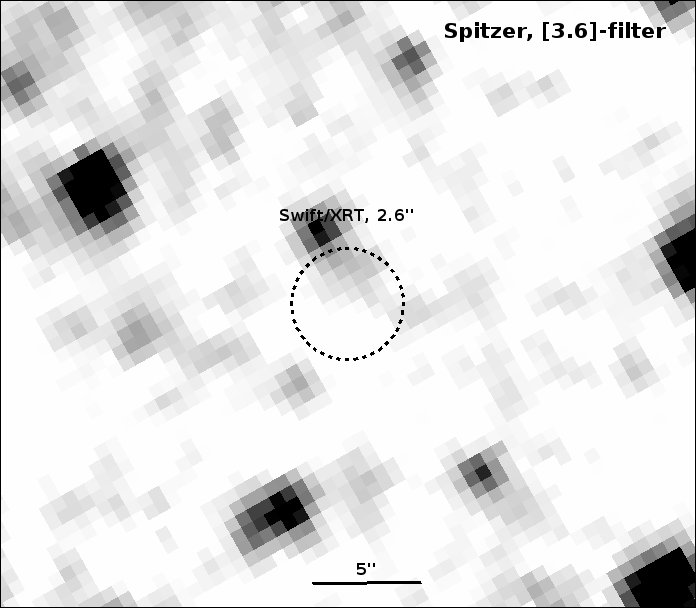}
\includegraphics[width=0.95\columnwidth,trim={1cm 7cm 0cm 3cm},clip]{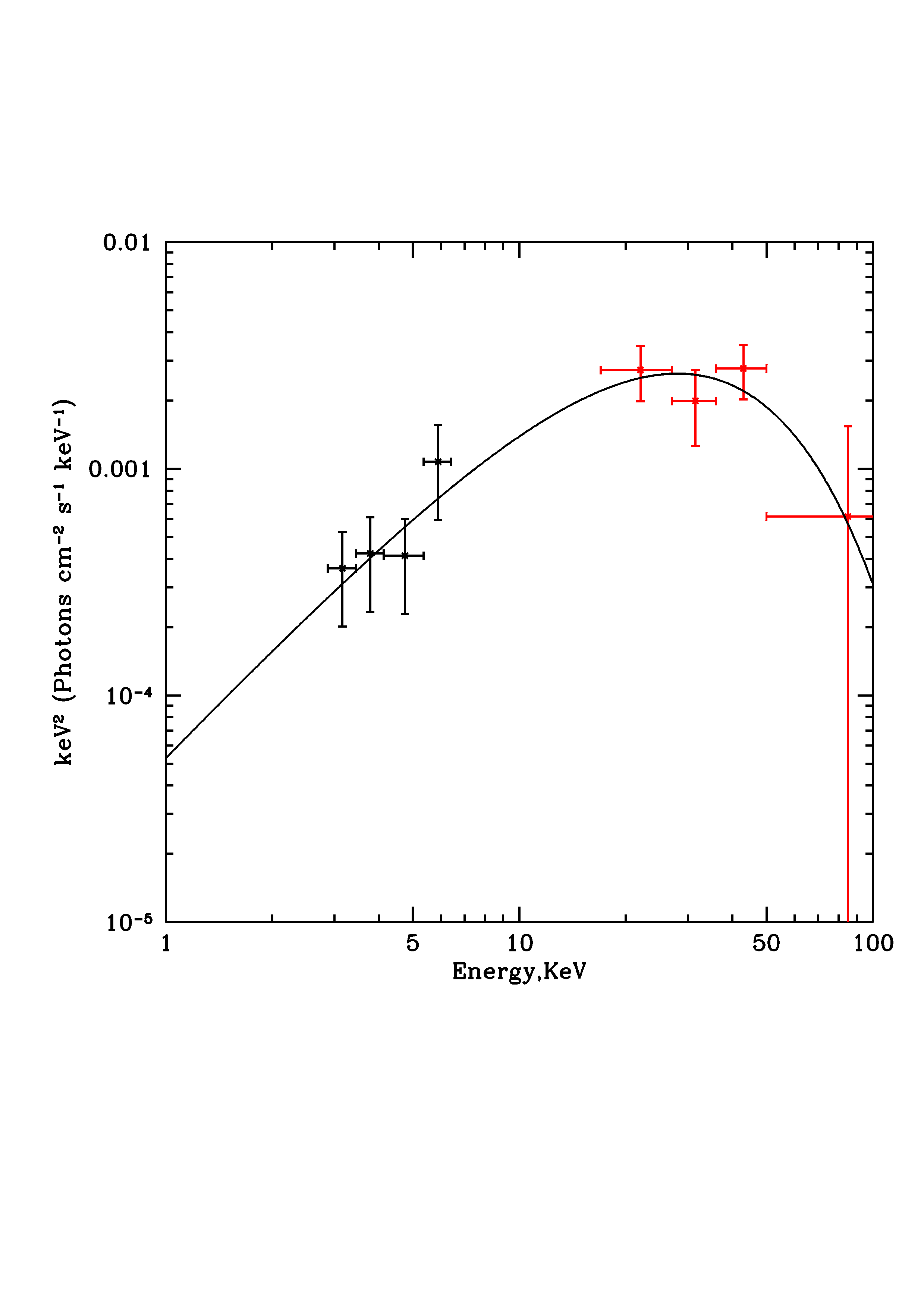}
\caption{ (Top) UKIDSS and GLIMPSE images of the sky region containing IGR\,J18044-1829. The dashed circles mark the XRT/Swift localization of the soft X-ray source (Table 1). (Bottom) The X-ray spectrum of IGR\,J18044-1829 reconstructed from the XRT/Swift and IBIS/INTEGRAL data. The solid line indicates the fit to the spectrum by a power law with and exponential cutoff. }  \label{fig:IGR8}
\end{figure*}

\begin{figure*}[!t]
\centering
\includegraphics[width=0.82\columnwidth,trim={0cm 0cm -0.5cm 0cm},clip]{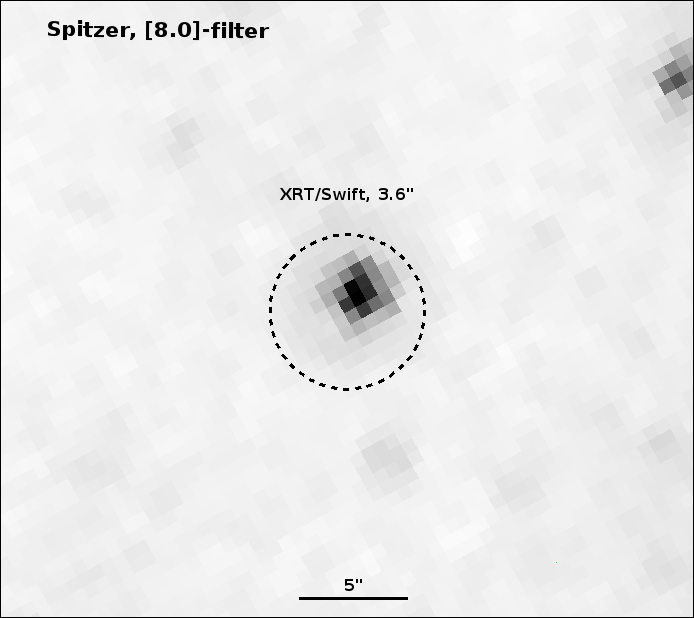}
\includegraphics[width=0.82\columnwidth,trim={-0.5cm 0 0cm 0cm},clip]{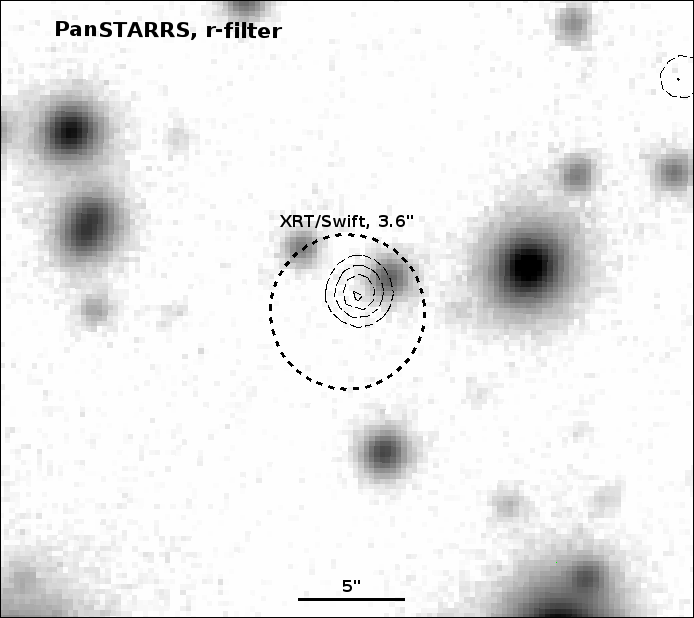}
\includegraphics[width=0.82\columnwidth,trim={0cm 0 0cm 0cm},clip]{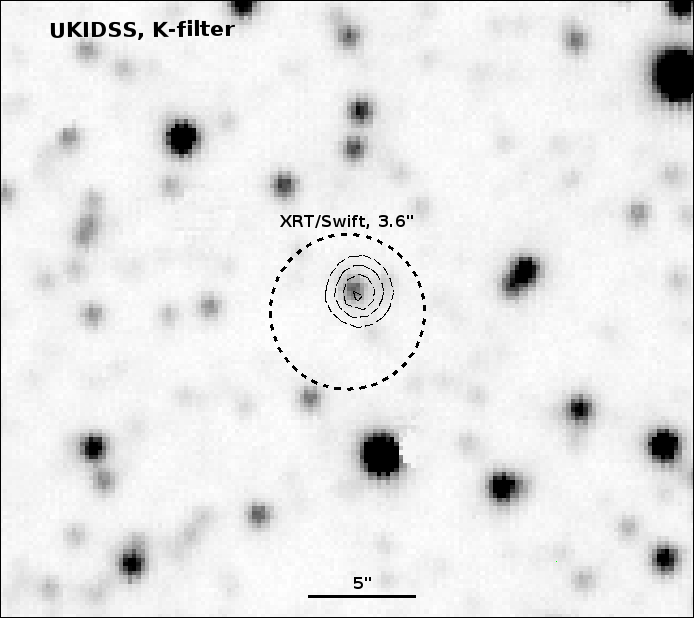}
\includegraphics[width=0.85\columnwidth,trim={0cm 7cm 1cm 4cm},clip]{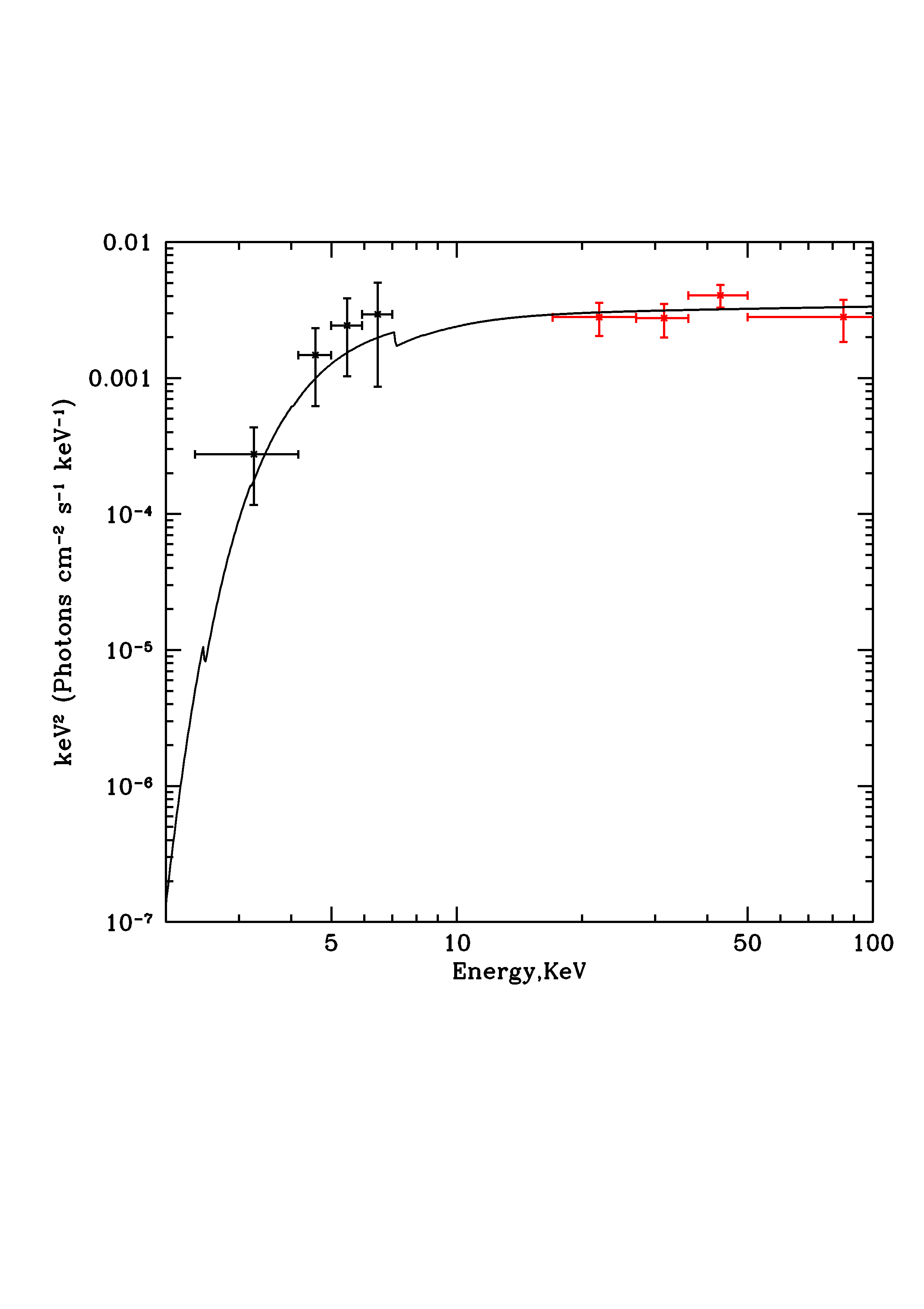}
\caption{GLIMPSE, PanSTARRS, and UKIDSS images of the sky region containing IGR\,J18141-1823. The dashed circles mark the XRT/Swift localization of the soft X-ray source (Table 1). The contours on the images in the r and K filters mark the position of the bright source in the mid-infrared (see the Spitzer image) falling into the XRT error circle. (Bottom right) The X-ray spectrum of IGR\,J18141-1823 from the XRT/Swift and IBIS/INTEGRAL data. The solid line indicates the fit to the spectrum by a power law with absorption.}  \label{fig:IGR7}
\end{figure*}

\begin{figure*}[!t]
\centering
\includegraphics[width=0.99\columnwidth,trim={1cm 0cm 0cm 1cm},clip]{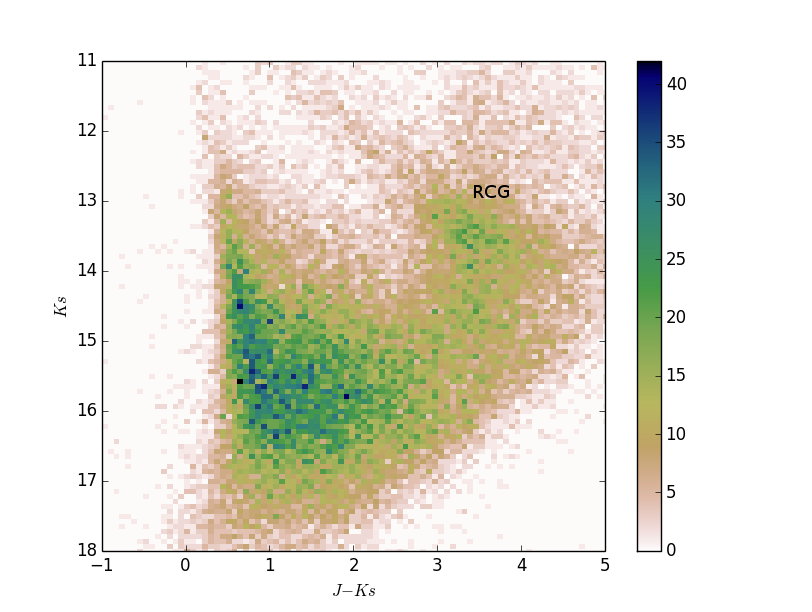}
\includegraphics[width=0.99\columnwidth,trim={1cm 0cm 0cm 1cm},clip]{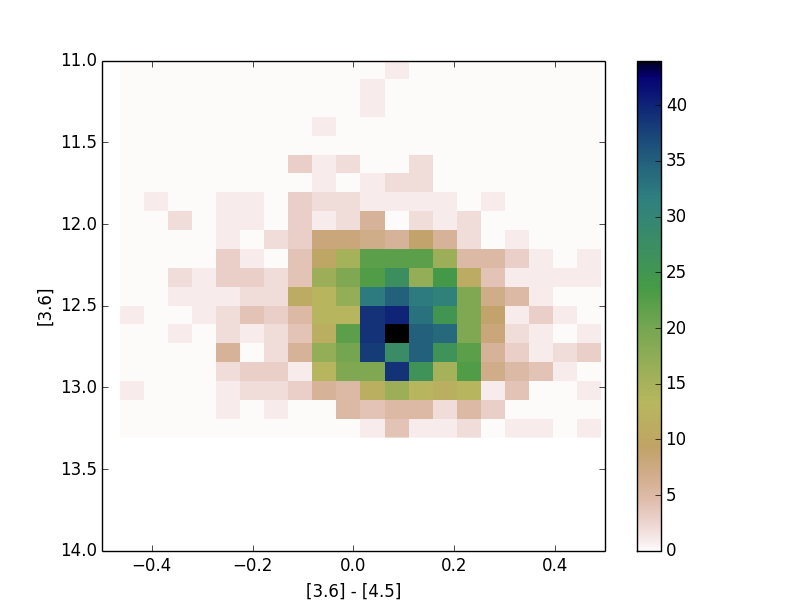}
\caption{Color–apparent magnitude diagrams constructed in different infrared filters for all stars within $10'$ of the source being investigated (from UKIDSS data (left)); only for the RCGs around the source being investigated distinguished from the total number of stars according to the algorithm described in the text (from Spitzer data (right)). The color gradients mark the number of stars in each bin of the corresponding histogram.}
\label{fig:abs7}
\end{figure*}

\begin{figure*}[!t]
\centering
\includegraphics[width=0.85\columnwidth,trim={0cm 0cm -1cm 0cm},clip]{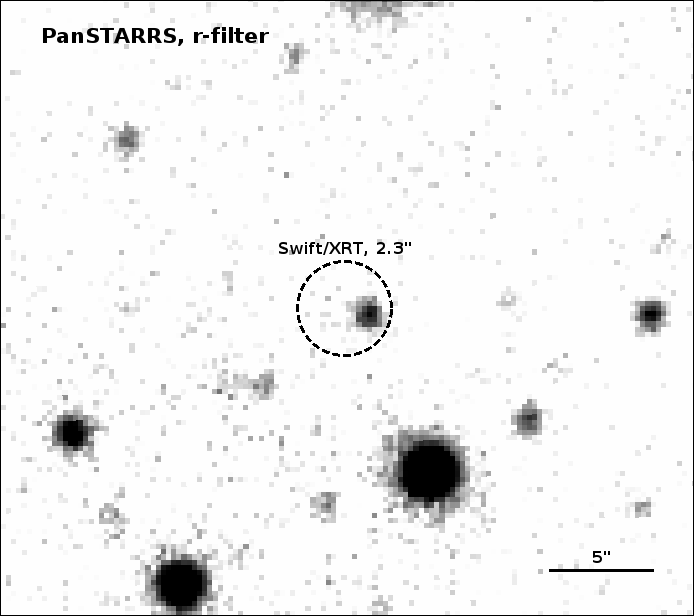}
\includegraphics[width=0.85\columnwidth,trim={-1cm 0 0cm 0cm},clip]{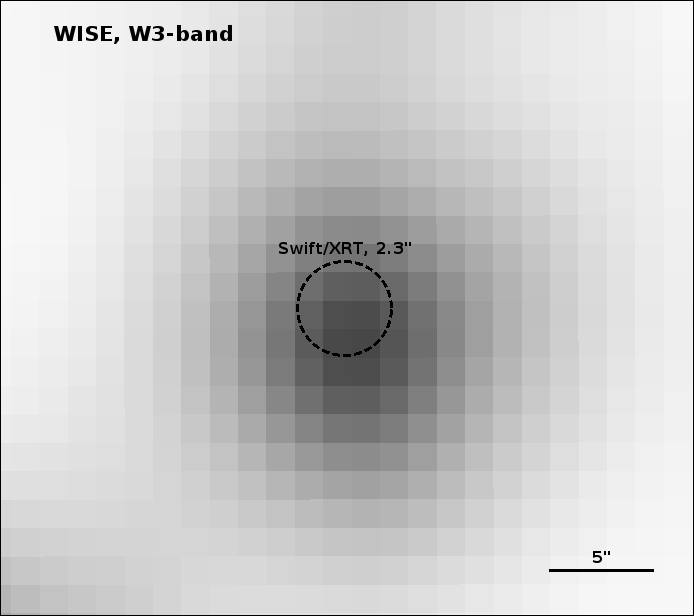}
\includegraphics[width=0.95\columnwidth,trim={1cm 7cm 0cm 3cm},clip]{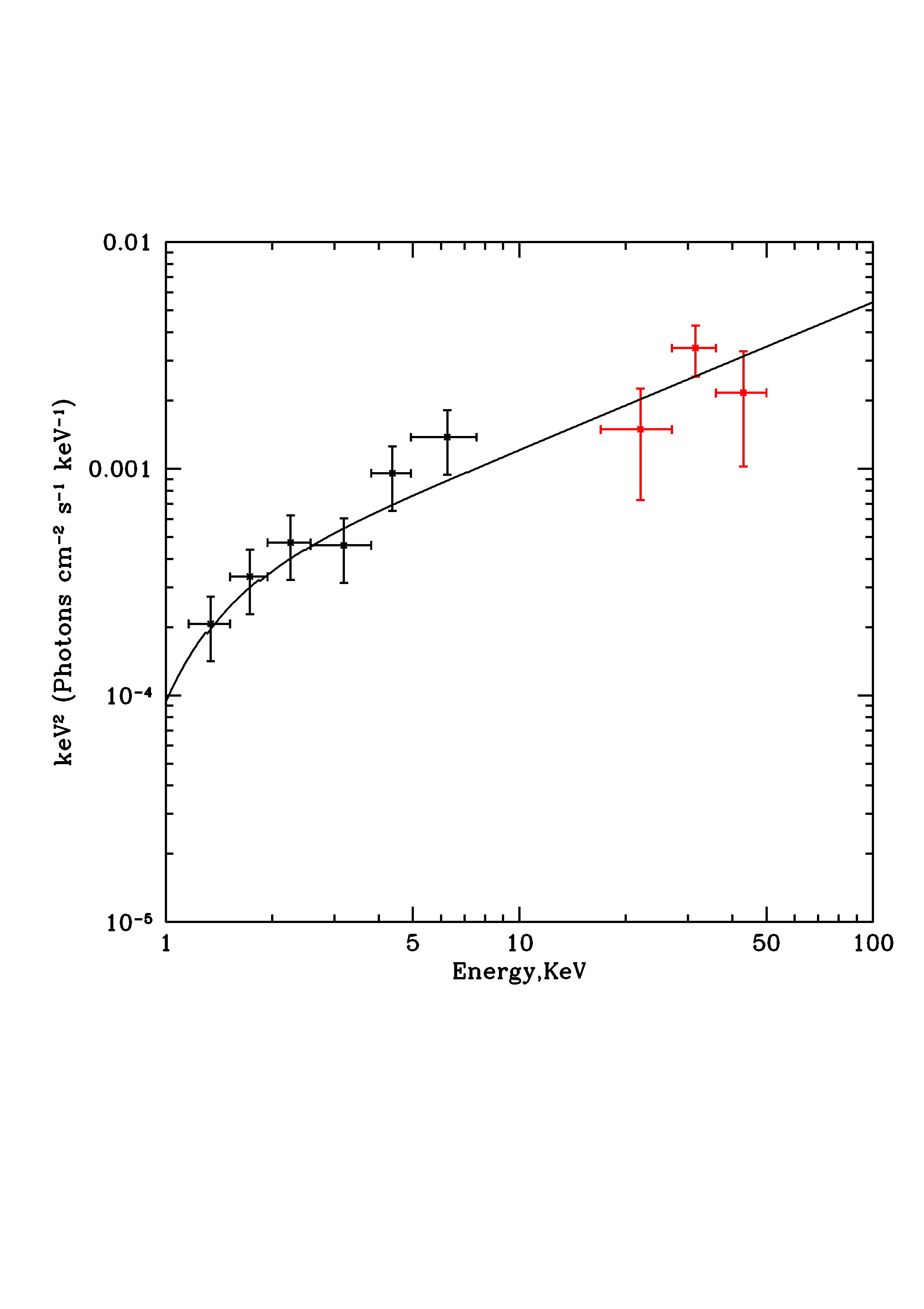}
\caption{(Top) PanSTARRS and WISE images of the sky region containing IGR\,J18544+0839 in the r and (left) and W3 (right) filters, respectively. The dashed circles mark the XRT/Swift localization of the soft X-ray source (Table 1). (Bottom) The X-ray spectrum of IGR\,J18544+0839 reconstructed from the XRT/Swift and IBIS/INTEGRAL data. The solid line indicates the fit to the spectrum by a power law with absorption. }  \label{fig:IGR10}
\end{figure*}

\subsection{IGR\,J18141-1823}
Krivonos et al. (2017) pointed out that IGR\,J18141-1823 most likely corresponds to 4PBC\,J1814.1-1822 from the Swift/BAT catalog. Using the publicly accessible XRT/Swift observations of the sky region around the source (ObsID 00043949001), we improved the localization of the object, determined its flux, and reconstructed its broadband X-ray spectrum by combining them with the IBIS/INTEGRAL data (see Fig. 10, Tables 1 and 2). The source’s spectrum can be described by a power law with a slope $\Gamma \simeq 1.95$ supplemented by interstellar absorption. 
The soft X-ray source falling into the INTEGRAL error circle is also known under the name 1SXPS\,J181414.8-182310 (Evans et al. 2014).
To determine a possible counterpart of the source and its nature, we performed a further study of the object’s fields in the mid-infrared using the archival data of the Spitzer space observatory (see Fig. 10). According to the data from the GLIMPSE/Spitzer sky survey, an infrared source with colors $([3.6 ] - [4.5]) = 0.874$ and $([5.8] - [8.0]) = 0.684$ (without absorption) is close to the center of the error circle of the soft X-ray source that we associate with IGR\,J18141-1823. According to the color classification in Stern et al. (2012, $([5.8] - [8.0]) > 0.6; ([3.6] - [4.5]) > 0.2\times([5.8] - [8.0]) + 0.18; ([3.6] - [4.5]) > 2.5\times ([5.8] - [8.0]) - 3.5$; see Fig. 4 in this paper), such colors correspond to an AGN with broad emission lines. However, the above values disregard the influence of interstellar extinction on the source’s colors, whereas the above criteria are valid for the true colors of objects. The mid-infrared emission is known to be affected weakly by extinction. However, since the source being investigated is close to the Galactic plane ($l\approx 12.4^{\circ}$, $b\approx -0.42^{\circ}$), this question requires a separate consideration. A significant advantage from the viewpoint of the position in the sky is the closeness of IGR\,J18141-1823 to the Galactic bulge, because this allows the above estimates to be made in a fairly understandable way.
As a reasonable approximation we assume that for sky regions toward the Galactic bulge close to the
Galactic plane the distribution of absorbing matter relative to the central part is nearly symmetric. Thus, in what follows, we will take the extinction and reddening to the far edge of the Galaxy to be twice those measured to the Galactic bulge in the corresponding directions. We consider the reddenings obtained in this way as the maximum possible ones in the Galaxy for a given direction, i.e., this correction is suitable precisely for extragalactic objects, whereas for Galactic ones it must be smaller.
First of all, let us determine the reddening to the Galactic bulge $E([3.6] - [4.5])_{bulge}$. This can be done by investigating the positions of red clump giants
(RCGs) on the color-magnitude $[3.6] ([3.6] - [4.5])$ diagram constructed in the corresponding direction for all star-like objects within $10\arcmin$ around the source being investigated. Such a significant radius was chosen, because on the edge of the bulge the RCG density is much lower than that at its center. We then compare the observed RCG color with the true RCG color, which is $([3.6] - [4.5])_0 = -0.10\pm 0.02$ (Karasev et al. 2018) for the bulge, and, as a result, we obtain the sough-for reddening $E([3.6] - [4.5])_{bulge}$.
However, when investigating the color-apparent magnitude diagram constructed for mid-infrared filters, it is very difficult to determine the RCG position on it directly (by investigating the local density of objects on the diagram and the broadening of the giant branch). The point is that in these colors objects of different classes differ little between themselves. Moreover, RCGs are usually located at magnitudes close to the detection limits of the corresponding surveys. To solve this problem, we use the approach applied in Karasev et al. (2018): first we find RCGs on the diagrams constructed for all star-like objects of the sky region being investigated in near-infrared filters (for example, Ks(J - Ks)) from the VVV or UKIDSS data; then we find a correspondence for these RCGs in the mid-infrared. Thus, having identi- fied RCGs on the $[3.6] ([3.6] - [4.5])$ diagram, we can determine their apparent color (Fig. 11).
As a result of the above actions, we obtained the reddening to the bulge $E([3.6] - [4.5])_{bulge} = 0.20\pm 0.03$. In accordance with the above assumption about a symmetric distribution of absorbing matter, the total reddening in the Galaxy toward the source is $E([3.6] - [4.5])_{Gal} \simeq 0.4$. As a result, the color of the infrared counterpart of the source being investigated corrected for the total reddening is $([3.6] - [4.5])_0 \simeq 0.47$. We did not additionally investigate the reddening of the source’s color $([5.8] - [8.0])$ due to its smallness even for the sky regions toward the Galactic plane. Thus, even applying the maximum possible correction to the counterpart’s colors, we find that it is most likely an extragalactic object (AGN). Nevertheless, additional infrared spectroscopic observations are needed to draw final conclusions about the nature of the source.

\subsection{IGR\,J18544+0839}

The XRT/Swift (ObsID 00010173001, June 2017) and IBIS/INTEGRAL observations of the source allowed its X-ray spectrum to be constructed in a wide energy range, which is described by a power law with photoabsorption (see Fig. 12, Tables 1 and 2). According to the PanSTARRS data, only one fairly bright object falls into the XRT error circle.

Using the WISE data, we can identify the PanSTARRS source with WISE J185422.29+083846.2, whose W2 magnitude is $W2 = 11.264\pm0.028$ and colors $W1-W2 = 1$ and $W2-W3 = 2.6$. According to the criteria($W1-W2>0.8$ and $W2 < 15.05$) from Stern et al. (2012), this points to an extragalactic nature of the counterpart (note also a significant brightness of the source in the reddest
WISE filter: $W4 = 5.958\pm0.061$) and allows the objects to be provisionally classified as an AGN.

Since IGR\,J18544+0839 is localized not toward the Galactic bulge, the reddening corrections for the above colors were taken from the standard maps obtained by Schlafly and Finkbeiner (2011) and the extinction law from Indebetouw et al. (2005). The corrections are small $(E(W1-W2) = 0.04)$ even for the first two filters and do not modify our preliminary conclusions about the nature of the source being investigated. Note also that the object’s color $(W2-W3)\simeq2.65$, in addition to the above criterion, also suggests that IGR\,J18544+0839 is most likely an AGN (according to the criteria from Yan et al. (2013)).

Note that spectroscopic observations are needed for the final confirmation of an extragalactic nature of the object and its redshift determination.

\newpage
\bigskip

\begin{table*}
\centering
\footnotesize{
   \caption{Parameters of the models for the spectra of the sources and their unabsorbed and redshift-corrected luminosities}
   \begin{tabular}{c|c|c|c|c|c|c}
     \hline
     \hline
           &       & &  & Normalizing
     &      \\
          Name    &  $\Gamma$ & $E_{cutoff}$, keV & $n_H, \times 10^{22}$ cm$^{-2}$ & relative & $log (L_{2-10,unabs})$, [erg s$^{-1}$]&\\
              &         &         &      & constant &  &\\
     \hline

     IGR\,J01017+6519    &  $1.24\pm0.12$ &  -- & $1.45\pm0.92$& 1 &$43.43\pm 0.05$ \\     
     IGR\,J08321-1808    & $ 1.05\pm0.21$ &  -- & -- & 1 & $44.03\pm0.09$  \\       
     
     IGR\,J11299-6557 &$0.84\pm 0.22$& $22.14\pm  9.72$ & -- & 1 & -- \\
                                     &$1.62\pm0.12$&           --     &$0.56\pm0.22$ & 1 & -- \\

      IGR\,J14417-5533 &  $1.82\pm0.08$ & -- & $0.65\pm0.12$ & 1 & -- \\
     
     IGR\,J16494-1740 &$0.55\pm0.21$& $30.02\pm11.86$ & -- & 1 & $42.58\pm 0.04 $\\
                                     &$1.23\pm0.10$&    --     &$1.16\pm0.75$ & 1 & $42.65\pm 0.04$\\

     IGR\,J17098-2344	 & $1.46\pm0.07$ & $27.10\pm9.81$ &  $0.13\pm0.02$& $0.53\pm 0.16$ & $43.13\pm0.01  ?$\\  
    
     IGR\,J17422-2108	 &$ 1.70\pm0.05$&  -- & $0.43\pm 0.02$ & $1.49\pm0.44$ & $43.86\pm 0.01 $ \\			
      
    IGR\,J18044-1829 &  $0.35\pm  0.46 $& $16.97\pm 8.33$& -- & 1 &--\\	
  
    IGR\,J18141-1823 &  $1.95\pm0.25 $& -- &   $23.15\pm 11.20$& 1 & --   \\				
   			
     IGR\,J18544+0839 &  $1.35\pm0.14$ & -- &$0.44\pm 0.45$ & 1 & -- \\

     \hline
    \end{tabular}
    }
\end{table*}

\begin{table*}
\centering
\footnotesize{
   \caption{Optical and infrared counterparts of the hard X-ray sources }
   \begin{tabular}{c|c|c|c|c|c}
     \hline
     \hline
           &       &      &      &  & \\
         Name     & RA (J2000) & Dec (J2000) & Z & Class of object &Notes\\
              &         &         &      &  &\\

     \hline

     IGR\,J01017+6519    &  01$^h$ 01$^m$ 58$^s$.26 &  65\fdg 21\arcmin 17\arcsec.87 & $0.085$ & Sy1 & WISE\,J010158.29+652117.5\\
     IGR\,J08215-1320    &  08$^h$ 21$^m$ 33$^s$.51 &  -13\fdg 21\arcmin 04\arcsec.51 & 0.014 & Sy2 &MCG-02-22-003 \\
     IGR\,J08321-1808    &  08$^h$ 31$^m$ 58$^s$.37 &  -18\fdg 08\arcmin 35\arcsec.21 & $0.135$ & Sy1& WISE\,J083158.37-180835.2 \\    
           IGR\,J11299-6557 &  11$^h$ 29$^m$ 56$^s$.444 & $-65$\fdg$ 55$\arcmin$ 21$\arcsec$.83$ & -- & AGN?& WISE J112956.44-655521.8\\

     IGR\,J14417-5533 &  14$^h$ 41$^m$ 18$^s$.742 & $-55$\fdg$ 33$\arcmin$ 35$\arcsec$.17$ & -- & AGN? & WISE J144118.74-553335.1 \\
     
     IGR\,J16494-1740    &  16$^h$ 49$^m$ 21$^s$.04 &  -17\fdg 38\arcmin 40\arcsec.17 & $0.027$ & Sy2 & ESO 586-4\\
     IGR\,J17098-2344	 &  17$^h$ 09$^m$ 44$^s$.70 &  -23\fdg 46\arcmin 53\arcsec.21 & 0.036 & NLSy1 or Sy1.2 ? &  WISE\,J170944.70-234653.1 \\  
     IGR\,J17422-2108	 &  17$^h$ 42$^m$ 11$^s$.43 &  -21\fdg 03\arcmin 53\arcsec.48 & 0.106 & Sy1& PSO\,J174211.443-210353.335 \\			
    IGR\,J18044-1829 &  18$^h$ 04$^m$ 34$^s$.042 & -18\fdg 30\arcmin 05\arcsec.09 & -- & ? &--\\	

     IGR\,J18141-1823 &  18$^h$ 14$^m$ 14$^s$.526 & -18\fdg 23\arcmin 10\arcsec.42 & -- & AGN? & G012.4048-00.4216\\				
   			
     IGR\,J18544+0839 &  18$^h$ 54$^m$ 22$^s$.297 & +8\fdg 38\arcmin 46\arcsec.22 & -- & AGN? & WISE\,J185422.29+083846.2\\

     \hline
    \end{tabular}
    }
\end{table*}

\section*{CONCLUSIONS}
We made optical identifications of 11 hard X-ray sources from the 14-year INTEGRAL all-sky survey. Six of them were shown to be Seyfert 1 and 2 galaxies. The redshifts were measured for them. According to a number of indirect selection criteria in the mid-infrared, four more objects can also be AGNs. However, infrared spectroscopic observations are needed for this to be finally confirmed. For one more object we failed to make an unequivocal optical identification due to the insufficient (for a crowded field) XRT/Swift localization accuracy. Note also that in most cases the X-ray data in different bands obtained at different times by different observatories agree satisfactorily and often do not require the introduction of an additional normalization factor or require the introduction of a sufficiently small one.

The results obtained here are summarized in Tables 2 and 3.

In conclusion, it should be noted that since the INTEGRAL position error circle for X-ray objects is fairly large and since many of the X-ray sources are transient, situations where the sources detected by the Swift and XMM-Newton observatories within the INTEGRAL error circle may turn out to be there serendipitously and may not correspond to a real hard X-ray source (which was inactive at the time of observation) are quite probable. Additional observations by the Nustar observatory with a high sensitivity in a wide energy range can give the necessary information to answer this question.

\vspace{15mm}
\section*{ACKNOWLEDGMENTS}
This work was supported by RSF grant no. ${14-22-00271}$. We are grateful to S.A. Grebenev for several valuable discussions and remarks that significantly improved this paper. We also thank the TUBITAK National Observatory (Turkey), the Space Research Institute of the Russian Academy of Sciences, and the Kazan State University for the support in using the Russian-Turkish 1.5-m telescope (RTT-150). M.V. Eselevich is grateful to Basic Research Program P-7 (“Transient and explosive processes in astrophysics”) of the Presidium of the Russian Academy of Sciences for partial support of the observations at the 1.6-m Sayan observatory telescope, which is part of the “Angara” center. We performed part of the work based on data from the GPS sky survey of the UKIRT telescope. This publication also uses data from the WISE observatory, which is a joint projection of the California Institute, Los Angeles, and the Jet Propulsion Laboratory/California Institute of Technology financed by NASA. We used data from the Spitzer space telescope, which is maintained by the Jet Propulsion Laboratory of the California Institute of Technology and NASA.

\section*{REFERENCES}

1. V. Afanasiev, S. Dodonov, V. Amirkhanyan, and A. Moiseev, Astrophys. Bull. {\bf 71}, 479 (2016).\\

2. I. Bikmaev, M. Revnivtsev, R. Burenin, and R. Syunyaev, Astron. Lett. {\bf 32}, 588 (2006).\\

3. I. Bikmaev, R. Burenin, M. Revnivtsev, S. Yu. Sazonov, R. A. Sunyaev, M. N. Pavlinsky, and N. A. Sakhibullin, Astron. Lett. {\bf 34}, 653 (2008).\\

4. A. J. Bird, A. Bazzano, A. Malizia, M. Fiocchi, V. Sguera, L. Bassani, A. B. Hill, P. Ubertini, and C. Winkler, Astrophys. J. Suppl. Ser. {\bf 223}, 10 (2016).\\

5. R. Burenin, A. Meshcheryakov, M. Revnivtsev, S. Yu. Sazonov, I. F. Bikmaev, M. N. Pavlinsky, and R. A. Sunyaev, Astron. Lett. {\bf 34}, 367 (2008).\\

6. R.Burenin, I.Bikmaev, M.Revnivtsev, J.A.Tomsick, S. Yu. Sazonov, M. N. Pavlinsky, and R. A. Sunyaev, Astron. Lett. {\bf 35}, 71 (2009).\\

7. R. A. Burenin, A. L. Amvrosov, M. V. Eselevich, V. M. Grigor’ev, V. A. Aref’ev, V. S. Vorob’ev, A. A. Lutovinov, M. G. Revnivtsev, S. Yu. Sazonov, A. Yu. Tkachenko, G. A. Khorunzhev, A. L. Yaskovich, and M. N. Pavlinsky, Astron. Lett. {\bf 42}, 295 (2016).\\

8. E.Churazov, R.Sunyaev, S.Sazonov, M.Revnivtsev, and D. Varshalovich, Mon. Not. R. Astron. Soc. {\bf 357}, 1377 (2005).\\

9. E. Churazov, R. Sunyaev, J. Isern, et al., Nature (London, U.K.) {\bf 512}, 406 (2014).\\

10. F. Durret, Kenichi Wakamatsu, T. Nagayama, S. Adami, and A. Biviano, Astron. Astrophys. {\bf 583}, 11 (2015).\\

11. R. Edelson and M. Malkan, Astrophys. J. {\bf 751}, 52 (2012). \\

12. P. A. Evans, A. P. Beardmore, K. L. Page, J. P. Osborne, P. T. O’Brien, R. Willingale, R. L. C. Starling, D. N. Burrows, et al., Mon. Not. R. Astron. Soc. {\bf 397}, 1177 (2009).\\

13. P. A. Evans, J. P. Osborne, A. P. Beardmore, K. L. Page, R. Willingale, C. J. Mountford, C. Pagani, D. N. Burrows, et al., Astrophys. J. Suppl. Ser. {\bf 210}, 24 (2014). \\

14. M.R.Goad, L.G.Tyler, A.P.Beardmore, P.A.Evans, S. R. Rosen, J. P. Osborne, R. L. C. Starling, F. E. Marshall, et al., Astron. Astrophys. {\bf 476}, 1401 (2007). \\

15. R.W.Goodrich, Astrophys.J. {\bf 342}, 224 (1989). \\

16. S. A. Grebenev, A. A. Lutovinov, S. S. Tsygankov, and I. A. Mereminskiy, Mon. Not. R. Astron. Soc. {\bf 428}, 50 (2013). \\

17. T. Hasegawa, Ken-ichi Wakamatsu, M. Malkan, K. Sekiguchi, J. W. Menzies, Q. A. Parker, J. Jugaku, H. Karoji, and S. Okamura, Mon. Not. R. Astron. Soc. {\bf 316}, 326 (2000). \\

18. R. Indebetouw, J. S. Mathis, B. L. Babler, M. R. Meade, C. Watson, B. A. Whitney, M. J. Wolff, M. G. Wolfire, et al., Astrophys. J. {\bf 619}, 931 (2005). \\

19. D. H. Jones, M. A. Read, W. Saunders, M. Colless, T. Jarrett, Q. Parker, A. Fairall, T. Mauch, et al., Mon. Not. R. Astron. Soc. {\bf 399}, 683J (2009).\\

20. S.F.Kamus, S.A.Denisenko, N.A.Lipin, V.I.Tergoev, P. G. Papushev, S. A. Druzhinin, Yu. S. Karavaev, and Yu. M. Palachev, J. Opt. Technol. {\bf 69}, 674 (2002). \\

21. D. I. Karasev and A. A. Lutovinov, Astron. Lett. {\bf 44}, 220 (2018). \\

22. D. I. Karasev, A. A. Lutovinov, and R. A. Burenin, Mon. Not. R. Astron. Soc. Lett. {\bf 409}, L69 (2010).\\

23. X.P.Koenig and D.T.Leisawitz, Astrophys. J.791, {\bf 2} (2014). \\

24. S. Kouzuma and H. Yamaoka, Mon. Not. R. Astron. Soc. {\bf 405}, 2062 (2010).\\

25. R. Krivonos, M. Revnivtsev, A. Lutovinov, S. Sazonov, E. Churazov, and R. Sunyaev, Astron. Astrophys. {\bf 475}, 775 (2007).\\

26. R. Krivonos, M. Revnivtsev, S. Tsygankov, et al., Astron. Astrophys. {\bf 519}, A107 (2010). \\

27. R.Krivonos, S.Tsygankov, A.Lutovinov, M.Revnivtsev, E. Churazov, and R. Sunyaev, Astron. Astrophys.
{\bf 545}, 7 (2012). \\

28. R. Krivonos, S. Tsygankov, I. Mereminskiy, A. Lutovinov, S. Sazonov, and R. Sunyaev, Mon. Not. R.
Astron. Soc. {\bf 470}, 512 (2017).\\

29. A.Lutovinov, R.Burenin, S.Sazonov, M.Revnivtsev, A. Moiseev, and S. Dodonov, Astron. Telegram 2759,
1 (2010). \\

30. A.Malizia,L.Bassani,V.Sguera,etal.,Mon.Not.R. Astron. Soc. {\bf 408}, 975 (2010). \\

31. D.J.Marshall, A.C.Robin, C.Reyle, M.Schultheis and S. Picaud, Astrophys. J. {\bf 453}, 635 (2006). \\

32. G. Marton, L. V. Toth, R. Paladini, M. Kun, S. Zahorecz, P. McGehee, and C. Kiss, Mon. Not. R. Astron. Soc. {\bf 458}, 4 (2016). \\

33. N. Masetti, R. Landi, M. Pretorius, et al., Astron. Astrophys. {\bf 470}, 331 (2007). \\

34. N.Masetti, P.Parisi, E.Palazzi, etal., Astron. Astrophys. {\bf 519}, 96 (2010). \\

35. I. A. Mereminskiy, R. A. Krivonos, A. A. Lutovinov, S. Yu. Sazonov, M. G. Revnivtsev, and R. A. Sunyaev, Mon. Not. R. Astron. Soc. {\bf 459}, 140 (2016). \\

36. S.V.Molkov, A.M.Cherepashchuk, A.A.Lutovinov, M. G. Revnivtsev, K. A. Postnov, and R. A. Sunyaev,
Astron. Lett. {\bf 30}, 382 (2004). \\

37. D.E.Osterbrock and R.W.Pogge, Astrophys. J. {\bf 297} (1985).\\

38. Planck Collab., Astron. Astrophys. {\bf 594}, 63 (2016). \\

39. M. Revnivtsev, R. Sunyaev, D. Varshalovich, V. Zheleznyakov, A. Cherepashchuk, A. Lutovinov, E. Churazov, S. Grebenev, and M. Gilfanov, Astron. Lett. {\bf 30}, 534 (2004). \\

40. M. G. Revnivtsev, S. Yu. Sazonov, S. V. Molkov, A. A. Lutovinov, E. M. Churazov, and R. A. Sunyaev, Astron. Lett. {\bf 32}, 145 (2006). \\

41. A. F. Rojas, N. Masetti, D. Minniti, E. Jimenez-Bailon, V. Chavushyan, G. Hau, V. A. McBride, L. Bassani, et al., Astron. Astrophys. {\bf 602}, 124 (2017). \\

42. E. F. Schlafly and D. P. Finkbeiner, Astrophys. J. {\bf 737}, 103 (2011). \\

43. D. J. Schlegel, D. P. Finkbeiner, and M. Davis, Astrophys. J. {\bf 500}, 525 (1998). \\

44. N. J. Secrest, R. P. Dudik, B. N. Dorland, N. Zacharias, V. Makarov, A. Fey, J. Frouard, and C. Finch, Astrophys. J. S. {\bf 221}, 12 (2015).\\

45. D.Stern, R.J.Assef, D.J.Benford, A.Blain, R.Cutri, A. Dey, P. Eisenhardt, R. L. Griffith, et al., Astrophys. J. {\bf 753}, 30 (2012). \\

46. J.Tomsick, S.Chaty, J.Rodriguez, etal., Astrophys. J. {\bf 701}, 811 (2009). \\

47. J. A. Tomsick, R. Krivonos, Q. Wang, A. Bodaghee, S. Chaty, F. Rahoui, J. Rodriguez, and F. M. Fornasini, Astrophys. J. {\bf 816}, 14 (2016). \\

48. W. Voges, B. Aschenbach, Th. Boller, H. Brauninger, U. Briel, W. Burkert, K. Dennerl, J. Englhauser, et al., Astron. Astrophys. {\bf 349}, 389 (2003). \\

49. C. Winkler, T.J.-L. Courvoisier, G. di Cocco, N. Gehrels, A. Gimenez, S. Grebenev, W. Hermsen, J. M. Mas-Hesse, et al., Astron. Astrophys. {\bf 411}, L1 (2003). \\

50. L. Yan, E. Donoso, Chao-Wei Tsai, D. Stern, R. J. Assef, P. Eisenhardt, A. W. Blain, R. Cutri, et al., Astrophys. J. {\bf 145}, 3 (2013). \\

{\it Translated by V. Astakhov}

\end{document}